\documentclass[letterpaper,titlepage,11pt]{article}

\pdfoutput=1

\usepackage{amsmath,amssymb,amsthm,mathrsfs,bbm}
\usepackage{latexsym,amscd,amsbsy,amsfonts,dsfont}
\usepackage{graphicx}
\usepackage{xcolor}
\usepackage[
      colorlinks=true,
      linkcolor=blue,
      urlcolor=blue,
      filecolor=black,
      citecolor=red,
      pdfstartview=FitV,
      pdftitle={},
        pdfauthor={Marco Astorino, Roberto Emparan, Adriano Vigano},
        pdfsubject={},
        pdfkeywords={},
        pdfpagemode=None,
        bookmarksopen=true
      ]{hyperref}\usepackage{caption}

\usepackage[utf8]{inputenc}
\usepackage{multirow}
\usepackage{makecell}
\usepackage{tikz}

\newcommand{\beq}{\begin{equation}}
\newcommand{\eeq}{\end{equation}}
\newcommand{\beqa}{\begin{eqnarray}}
\newcommand{\eeqa}{\end{eqnarray}}
\newcommand{\bea}{\begin{eqnarray}}
\newcommand{\eea}{\end{eqnarray}}

\newcommand{\nn}{\nonumber}

\def\clock{{\count0=\time
           \divide\count0 60
           \ifnum\count0<10 0\fi\the\count0
           \multiply\count0 -60 \advance\count0 \time
           :\ifnum\count0<10 0\fi \the\count0
         }}
\newcommand{\timestamp}{{\small\vbox{\hbox{\tt\jobname.tex}
\hbox{\the\day/\the\month/\the\year, \clock}}}}

\numberwithin{equation}{section}

\begin{document} 

\begin{titlepage}

\begin{flushright}
IFUM-1094-FT
\end{flushright}

\begin{center}

\phantom{ }
\vspace{1.5cm}

{\bf \LARGE{Bubbles of nothing in binary black holes\\ \vspace{10pt}  and black rings, and viceversa}}

\vskip 1cm

{\bf Marco Astorino${}^{a}$, Roberto Emparan${}^{b,c}$, Adriano Vigan\`o${}^{a,d}$}
\vskip 0.6cm

\small{${}^{a}$\textit{Istituto Nazionale di Fisica Nucleare (INFN), Sezione di Milano}}\\
\small{\textit{Via Celoria 16, I-20133 Milano, Italy}}
\medskip

\small{${}^{b}$\textit{Instituci\'o Catalana de Recerca i Estudis Avançats (ICREA),}}\\
\small{\textit{Passeig Llu\'is Companys 23, E-08010
Barcelona, Spain}}
\medskip

${}^{c}$\textit{Departament de F\'isica Qu\`antica i Astrof\'isica, Institut de Ci\`encies del Cosmos,}\\ \textit{Universitat de
Barcelona, Mart\'i i Franqu\`es 1, E-08028 Barcelona, Spain}
\medskip

${}^{d}$\textit{Dipartimento di Fisica, Universit\`a degli Studi di Milano}\\
\small\textit{Via Celoria 16, I-20133 Milano, Italy}

\bigskip

\texttt{marco.astorino@gmail.com}, 
\texttt{emparan@ub.edu}, \texttt{adriano.vigano@unimi.it}

\end{center}
\vskip 1.cm
\centerline{\bf Abstract} \vskip 0.2cm

\noindent
We argue that expanding bubbles of nothing are a widespread feature of systems of black holes with multiple or non-spherical horizons, appearing as a limit of regions that are narrowly enclosed by the horizons. The bubble is a minimal cycle that links the Einstein-Rosen bridges in the system, and its expansion occurs through the familiar stretching of space in black hole interiors. We demonstrate this idea (which does not involve any Wick rotations) with explicit constructions in four and five dimensions. The geometries of expanding bubbles in these dimensions arise as a limit of, respectively, static black hole binaries and black rings.
The limit is such that the separation between the two black holes, or the inner hole of the black ring, becomes very small, and the horizons of the black holes correspond to acceleration horizons of the bubbles. We also explain how a five-dimensional black hole binary gives rise to a different type of expanding bubble. 
We then show that bubble spacetimes can host black hole binaries and black rings in static equilibrium, with their gravitational attraction being balanced against the background spacetime expansion. Similar constructions are expected in six or more dimensions, but most of these solutions can be obtained only numerically. Finally, we argue that the Nariai solution can be regarded as containing an expanding circular bubble of nothing.

\end{titlepage}

\setcounter{tocdepth}{2}


\newpage


\section{Introduction and Summary}\label{sec:Introduction}

Expanding bubbles of nothing are simple but surprising solutions of gravitational theories with compact dimensions \cite{Witten-81}. They provide channels for the non-perturbative decay of Kaluza-Klein vacua, but they are also interesting as simple time-dependent spacetimes that share many features with de Sitter cosmologies \cite{Aharony:2002cx}. This latter view, more than the former, will be relevant in this article, where we present a suggestive new way of regarding these bubbles, and investigate their relation to some black hole systems.\footnote{Other aspects of the relation between black holes and bubbles of nothing have been studied in \cite{emparan-weyl,horowitz-maeda,Elvang:2002br,Elvang:2004iz,Tomizawa:2007mz,Iguchi:2007xs,kastor-ray,Kunz:2008rs,Yazadjiev:2009gr,Nedkova:2010gn,Kunz:2013osa}.}

More specifically, we will explain that expanding bubbles of nothing are a pervasive feature of systems of black holes with multiple or non-spherical horizons.\footnote{The precise notion of the topology that is required will become clearer below.} To demonstrate the idea, we will show that expanding bubbles of nothing arise as a limit of static black hole binaries (in four dimensions) and of black rings (in five dimensions). These systems allow us to illustrate a general phenomenon using explicit exact solutions of vacuum gravity. We expect that versions of all the constructions are possible in six or more dimensions, but then the solutions must be obtained numerically. Other lesser-known kinds of bubbles, in five or more dimensions, arise from different black hole binaries and will be briefly examined. Towards the end of the article we will discuss more general configurations using topological arguments, and argue that expanding bubbles are also present in systems such as the Schwarzschild-de Sitter  and Nariai solutions.

We will also study how the expansion in bubble spacetimes acts on gravitationally interacting systems, in a manner similar to inflation in de Sitter. We will show that bubbles in four and five dimensions admit within them black hole binaries and black rings in static (although unstable) equilibrium, their attraction being balanced against the expansion of the background spacetime. The same mechanism is expected to work in more general situations for which exact solutions are not available.

\paragraph{Expanding bubbles of nothing from black hole binaries and black rings.}

The solution for an expanding bubble of nothing was originally presented in  \cite{Witten-81} in the form
\begin{equation}\label{bon1}
    ds^2=r^2\left( -dT^2 +\cosh^2 T\,d\Omega_n \right) +\frac{dr^2}{1-\frac{r_0^n}{r^n}}+\left(1-\frac{r_0^n}{r^n}\right)r_0^2\, d\phi^2\,.
\end{equation}
This is obtained from the Schwarzschild-Tangherlini solution in $n+3$ dimensions by rotating to imaginary values the time coordinate and one polar angle. However, the relation between black holes and bubbles that we will discuss is of a different kind and does not involve any such rotation. Since the coordinate $\phi$ must be periodically identified, $\phi\sim \phi + 4\pi/n$, the solution has Kaluza-Klein asymptotics, but the latter fact will also be of minor relevance for our discussion.

To understand the geometry, observe that the time-symmetric section at $T=0$ is the product of a `cigar' along the $(r,\phi)$ directions, and spheres $S^n$ of radius $r$. These spheres cannot be shrunk to zero size since they reach a minimum radius at $r=r_0$. The minimal sphere constitutes the bubble of nothing, and when it evolves for $T>0$, it expands in a de Sitter-like fashion. 

The coordinates in \eqref{bon1} cover the spacetime globally, but we can also write it using `static-patch' coordinates\footnote{The change is $t=\mathrm{arctanh}(\tanh T/\cos\chi)$, $\xi=\cosh T \sin\chi$, where $\chi$ is a polar angle of $S^n$.\label{foot:change}}, where
\begin{equation}\label{bonstat}
   ds^2=r^2\left( -(1-\xi^2)dt^2 +\frac{d\xi^2}{1-\xi^2}+\xi^2\,d\Omega_{n-1} \right) +\frac{dr^2}{1-\frac{r_0^n}{r^n}}+\left(1-\frac{r_0^n}{r^n}\right)r_0^2\, d\phi^2\,.
\end{equation}
We could set $\xi=\cos\theta$ to relate it more manifestly to the Schwarzschild-Tangherlini solution with imaginary $t$ and $\phi$, but the form above makes clearer the existence of a de Sitter-like horizon at $\xi^2=1$. In the full spacetime, this is an infinite acceleration horizon that extends from the bubble at $r=r_0$ to infinity. Observers who sit on the bubble midpoint between the horizons, that is, near $r=r_0$ and around $\xi=0$, find themselves partly surrounded (but not enclosed) by a horizon with topology $S^{n-1}\times \mathbb{R}^2$. Like in de Sitter, these observers do not have access to the entire $S^n$ bubble. They only see the half of it that remains static, while the portion of the bubble beyond the horizon expands exponentially.
\begin{figure}[t]
        \centering
         \includegraphics[width=\textwidth]{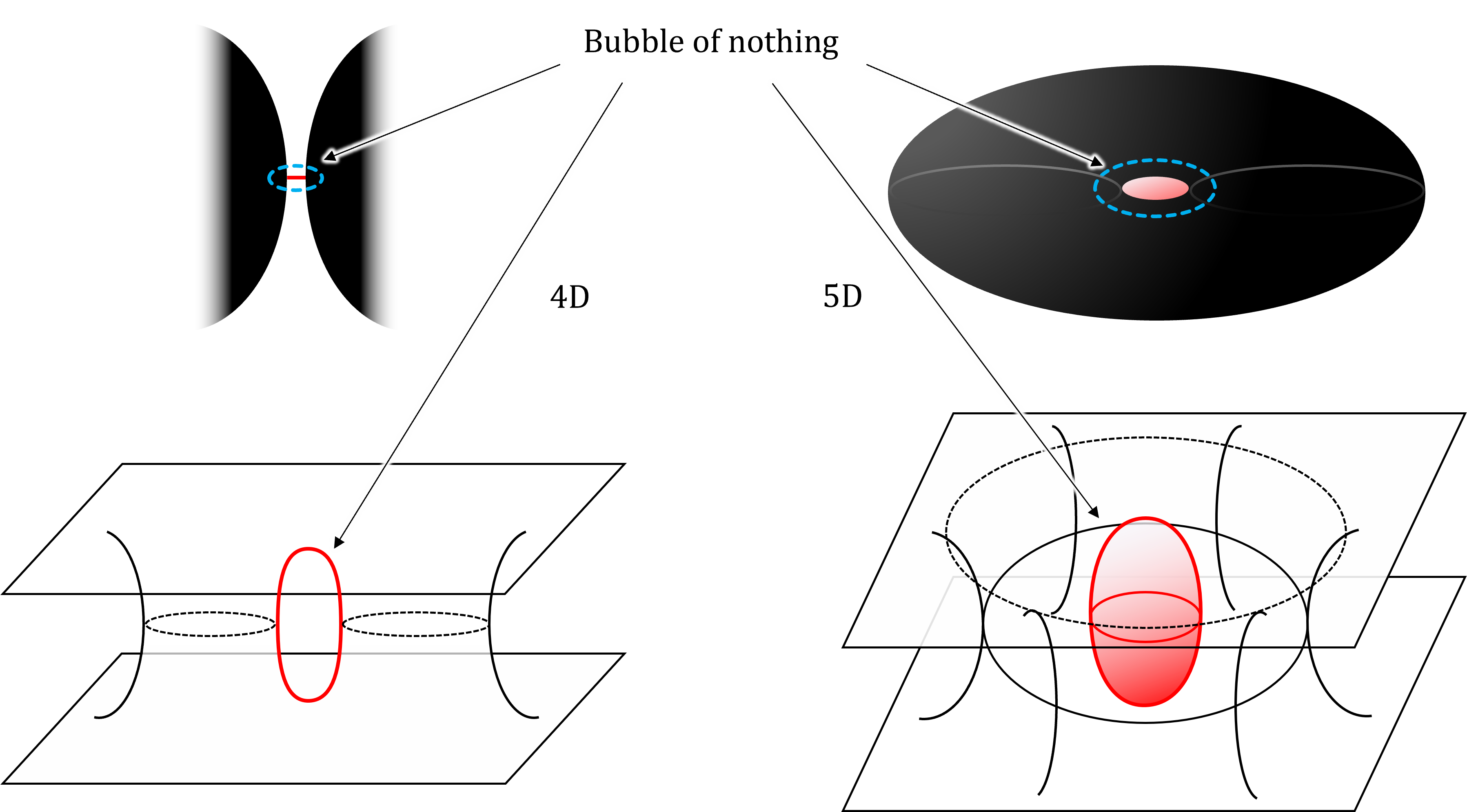}
        \caption{ \small Bubbles of nothing as limits of black hole systems. The top pictures are illustrative cartoons, and the bottom ones show time-symmetric spatial sections of the maximally extended solutions. Left top: The 4D bubble of nothing arises as the geometry in between two black holes, in the limit when their size is very large. The horizons of the black holes correspond to acceleration horizons of the bubble. Left bottom: The bubble is a minimal circle (in red) linking the Einstein-Rosen throats of the two black holes. This circle  encloses `nothing', and its expansion occurs as the throats stretch in the black hole interiors. The angle $\phi$ around the rotation axis is suppressed in these figures. Right top: the 5D bubble is similarly recovered from the central region of a very fat black ring. Right bottom: The bubble is a sphere (in red) that wraps the portion of the Einstein-Rosen bridge in the inner `hole' of the ring. In the bottom figure, the ring's $S^2$ is not represented. In the solutions we discuss, the black hole binary and the black ring are kept static by semi-infinite cosmic strings and cosmic membranes (not shown), respectively, which pull them outwards, but other means of maintaining them in equilibrium are possible.}
        \label{fig:bubblebhs}
\end{figure}

Let us examine the case $n=1$ of a four-dimensional expanding bubble. This is seldom considered when studying Kaluza-Klein spacetimes, but we will give it a new twist. The sphere $S^{n-1}$ now consists of the two endpoints of the interval $-1\leq \xi\leq 1$, so the observer in the bubble lies between two approximately planar (for $r\approx r_0$) acceleration horizons. Such Rindler-type horizons are known to describe the geometry near a black hole, and we will find that this interpretation is also apt for the bubble geometries \eqref{bonstat}. That is, we will show that the four-dimensional bubble appears as the geometry in between two black holes, when they are separated a distance much smaller than their radius (see Fig.~\ref{fig:bubblebhs}).

One may wonder in what sense can a black hole binary contain an expanding bubble. The answer is much the same as for the static-patch metric \eqref{bonstat}. When $n=1$, the bubble is a circle that links the Einstein-Rosen bridges of the black hole pair---a minimal cycle that encloses nothing (Fig.~\ref{fig:bubblebhs}, bottom left).\footnote{Strictly speaking, the cycle $\Omega_1$ in \eqref{bon1} need not be a compact $S^1$, but we will take it to be so. In the binary, we are identifying asymptotic regions to yield the smallest maximal analytic extension.} The static observer in between the two black holes is limited by the horizons to only have access to a portion of this circle, namely, the segment of the axis between the two horizons. The rest of the circle lies beyond the horizons. Initially, at $T=0$, this other half-circle is another segment between the two Einstein-Rosen throats. As $T$ evolves, these throats stretch, so the portions of the circle inside the black holes expand in time. The expansion of the bubble is then the familiar stretching that occurs in the interior of the black hole.\footnote{The same effect is responsible for the growth of holographic volume complexity \cite{Susskind:2014rva}.} The compactification of the $\phi$ direction is a consequence of focusing on a small region around the symmetry axis, so the radius of the $\phi$ circles can only reach a finite maximum.

Exact solutions for a static configuration of a pair of black holes, kept apart by semi-infinite cosmic strings that pull on them, have been known for long \cite{bach-weyl,israel-khan}. We will use them to explicitly exhibit the limit where they reduce to \eqref{bonstat}. We emphasize that there is no Wick-rotation involved in this connection: the time and angular coordinates retain their physical meaning throughout the limit.

The five-dimensional bubble, described originally in \cite{Witten-81}, also admits a similar interpretation. Now the acceleration horizon, with topology $S^1\times \mathbb{R}^2$, is connected. We will find that \eqref{bonstat} with $n=2$ arises as the limit of a black ring, with horizon topology $S^1\times S^2$, when the size of the $S^2$ is much larger than the inner rim of the ring circle. The static coordinates only cover the hemisphere of the $S^2$ bubble that consists of the disk of the inner `hole' of the ring. In global coordinates, the $S^2$ is a minimal sphere that wraps the Einstein-Rosen bridge in the inner hole of the ring. In this case, we will use the solution, first found in \cite{emparan-weyl}, for a static black ring held in place by an infinite cosmic membrane attached to the outer rim of the ring.
We expect that this construction generalizes to all $n\geq 3$, but the required solutions, with horizons of topology $S^{n-1}\times S^2$, are only known numerically \cite{Kleihaus:2009wh,Kleihaus:2010pr}.

We will also briefly discuss how a certain type of five-dimensional black hole binary, in the limit of small separation, gives rise to a five-dimensional expanding bubble of a different kind than the $n=2$ bubble above: the minimal cycle is not a single sphere $S^2$, but two $S^2$ that lie on orthogonal spaces and which touch each other at both north and south poles. They compactify the spacetime on a two-torus (instead of a circle). A more general discussion of the topology of other configurations will be presented in the concluding section.

Let us mention that the specific mechanism that keeps the black hole pair, or the black ring, in static equilibrium is not an essential aspect of the construction. Cosmic strings and membranes, in the form of conical deficits along the outer symmetry axes of the systems, are easy to work with, but the two black holes could also carry electric charges of opposite sign (a dihole \cite{Emparan:1999au,Emparan:2001bb}) and be held in equilibrium by an external electric field, namely, a fluxbrane. A similar construction is also possible for dipole black rings \cite{Emparan:2004wy}. As long as the black holes are not extremal, they will have bifurcation surfaces and there will be expanding bubbles, with the effects of the electric field becoming negligible in the region between the horizons, since the external fluxbrane polarizes the system so as to cancel the opposite fluxbrane-like field between the charged black holes \cite{Emparan:2001gm}. Other equilibration methods are possible, but their differences only show up far from the gap between the horizons. In the limit to the bubble solution, the distinctions between these geometries disappear.

Indeed, the explanation we have given should make clear---and we will elaborate further on this in the concluding section---that, as long as the black holes in a binary have bifurcate horizons, expanding bubbles are also present in them even if the horizons are not static but dynamically merge and collapse---but in these cases the expansion of the bubble only lasts a finite time.

\paragraph{Black hole binaries and black rings inside expanding bubbles.} The previous remarks bring us to the other main subject of this article, namely, the static equilibrium configurations of black hole binaries and black rings.
\begin{figure}[t]
        \centering
         \includegraphics[width=.8\textwidth]{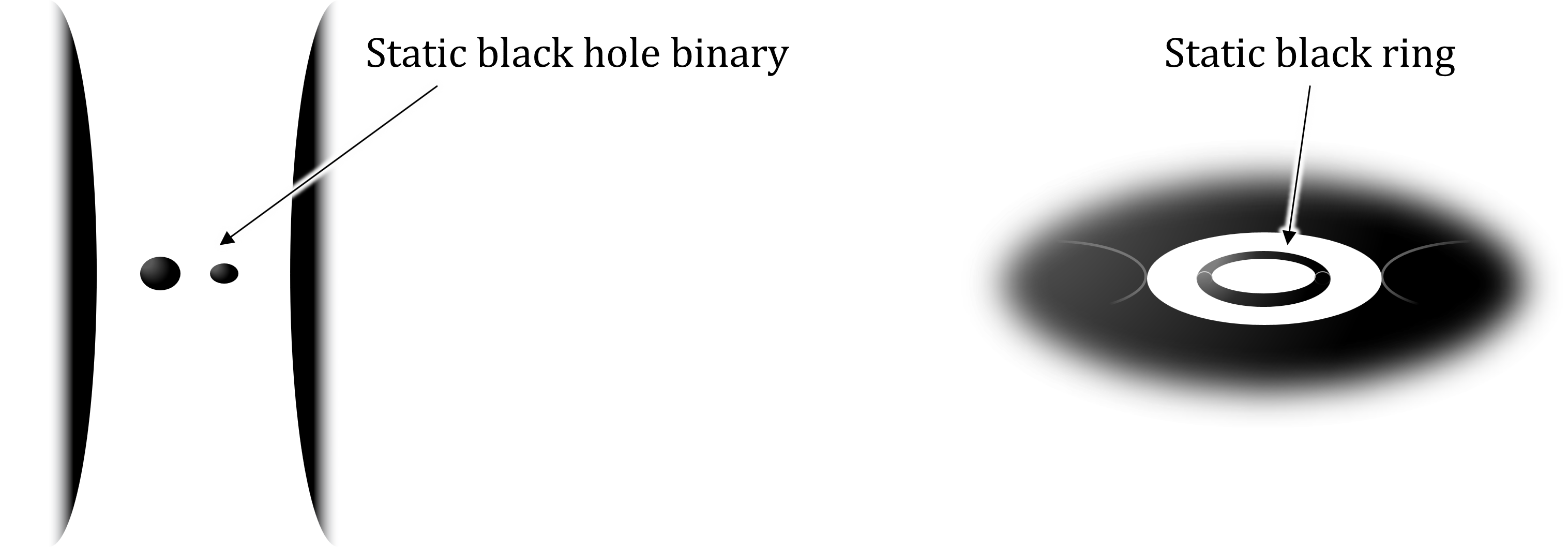}
        \caption{ \small Static black hole binaries and black rings obtained by placing them inside expanding bubbles of nothing, which surround them with acceleration horizons. The binary need not be symmetrical. These configurations can be regarded as limits of double nested static binaries (four black holes along a line), and of two coaxial black rings.}
        \label{fig:bhsinbubbles}
\end{figure}
Recently, an exact solution was presented where a static black hole binary is held in balance by the gravitational pull of distant masses away from the binary axis---these masses being modelled by multipolar source distributions at infinity \cite{Astorino:2021dju,charged-binary}. This is one of the few known exact solutions where a binary is maintained in equilibrium through purely gravitational fields.

Another possibility for balancing the attraction in the binary is the expansion of spacetime. Indeed, one expects that such binaries in (unstable) equilibrium exist in the de Sitter universe, but the solutions can only be constructed approximately for very small black holes, or numerically \cite{bensonetal}. Nevertheless, the expansion in a bubble should achieve the same effect. To prove this, we will construct exact solutions where an expanding bubble hosts a black hole binary (see figure \ref{fig:bhsinbubbles} left) (possibly with different masses) in static equilibrium.
The interpretation of the expanding bubbles given above provides another explanation for why this is possible: we can introduce a binary of two small black holes in the gap between two very large black holes, and then tune the distances between them so that the attraction in the small binary is balanced by the pull of the larger black holes.

The analogues of these configurations involving black rings in five dimensions can also be readily constructed  (see figure \ref{fig:bhsinbubbles} right). We will present a solution for a static black ring inside a five-dimensional bubble of nothing, and show that it can be recovered as the limit of a concentric, static double black ring system.

All these metrics can be given in exact closed form since they are Weyl solutions, which admit a systematic construction with an arbitrary number of collinear black holes, or concentric black rings \cite{Weyl:1917gp,bach-weyl,israel-khan,emparan-weyl}. The configurations are characterized by their rod structure \cite{emparan-weyl,Harmark:2004rm}, which specifies the sources along the different symmetry axes. This structure makes transparent the features of all the constructions discussed above and their limits. Indeed, the connections between the black hole binary and the static black ring, and the corresponding expanding bubbles in four and five dimensions, have been apparent at least since the analysis in \cite{emparan-weyl}. Nevertheless, to our knowledge, this connection does not appear to be widely known, and it has not been examined in detail in the literature.

\medskip

The remainder of the paper is organized as follows. In the next section we describe in detail how the solutions for the static black ring and static black hole binary have a limit to expanding bubbles of nothing. We also introduce Weyl metrics and discuss how these limits become very transparent in terms of rod structures. Other, less familiar binaries and bubbles in five dimensions are briefly discussed. In Sec.~\ref{sec:bhsfrombubbles} we construct static black holes binaries and black rings inside bubbles of nothing, and show that equilibrium can be achieved by appropriately tuning the parameters in the configurations. In Sec.~\ref{sec:otherconfig} we provide a taste of the many possible configurations and limits that are afforded by generalizing the Weyl constructions in this article. We conclude in Sec.~\ref{sec:discuss} emphasizing the wide scope and generality (and some limitations) of the constructions we have uncovered, and mentioning other instances of geometries that contain expanding bubbles of nothing, such as the Nariai solution.

\section{Bubbles as limits of black hole binaries and black rings}

We will now show explicitly how the metrics for the bubbles of nothing in 5D and 4D are recovered as limits of static black ring and binary black hole solutions in the manner illustrated in Fig.~\ref{fig:bubblebhs}.

\subsection{From black ring to bubble}\label{subsec:ringtobubble}

The simplest instance is the relationship between the five-dimensional bubble of nothing of~\cite{Witten-81} and the static black ring of~\cite{emparan-weyl}.
The metric of the latter is\footnote{In appendix~\ref{app:otherring} we do this study in the coordinates used in~\cite{Emparan:2004wy}.}
\begin{equation}
\label{bring}
ds^2=-\frac{F(x)}{F(y)}d\tilde t^2
+\frac{R^2}{(x-y)^2}\biggl[ F(x)\biggl( (y^2-1)d\tilde\psi^2 + \frac{F(y)}{y^2-1}dy^2\biggr)
+F(y)^2\biggl( \frac{dx^2}{1-x^2}+\frac{1-x^2}{F(x)}d\tilde\phi^2\biggr)\biggr] \,,
\end{equation}
with
\begin{equation}\label{Fring}
F(\xi)=1-\mu \xi\,.
\end{equation}
Readers unfamiliar with these $(x,y)$ coordinates are referred to~\cite{emparan-weyl} and~\cite{Emparan:2006mm} for a detailed explanation.
Roughly, $x\in[-1,1]$ is the cosine of the polar angle of the ring's $S^2$, and $-1/y\in (0,-1/x)$ is a radial coordinate away from these spheres.
The coordinates $\tilde\psi$ and $\tilde\phi$ are, respectively, the angle of the $S^1$ and the azimuthal angle of the $S^2$ of the black ring.
The parameter $R$ sets the scale for the size of the black ring, and varying $\mu\in[0,1)$ changes its shape from thin to fat.
The horizon lies at $y=-\infty$, and the absence of conical singularities along the $\tilde\psi$ rotation axis at $y=-1$, and in the inner disk of the ring at $x=1$, is obtained when we identify
\begin{equation}
\tilde\psi\sim \tilde\psi +2\pi\sqrt{1+\mu} \,, \qquad
\tilde\phi\sim \tilde\phi +2\pi\sqrt{1-\mu} \,.
\end{equation}
In order to make the ring very big and fat, and to blow up the inner disk region, we will take $\mu\to 1$ and $R\to\infty$ while zooming in onto $x\approx 1$. For this purpose, we change
\begin{equation}
x=1-\frac{r^2-r_0^2}{2R^2} \,, \qquad
\mu= 1-\frac{r_0^2}{2R^2} \,,
\end{equation}
where $r$ and $r_0$ are a new coordinate and a constant parameter that remain finite as $R\to\infty$.
In addition we introduce a coordinate $\xi$ via
\begin{equation}
y=-\frac{1+\xi^2}{1-\xi^2} \,,
\end{equation}
and rescale the Killing coordinates to have canonical normalization,
\begin{equation}
\tilde t=2 R t\,, \qquad
\tilde\psi =\sqrt{2}\psi\,, \qquad
\tilde\phi=\frac{r_0}{\sqrt{2} R}\phi\,.
\end{equation}
Then, in the limit $R\to\infty$, the metric~\eqref{bring} becomes
\begin{equation}\label{5dbubble}
ds^2\to r^2\biggl( -(1-\xi^2)dt^2 +\frac{d\xi^2}{1-\xi^2}+\xi^2\,d\psi^2 \biggr) +\frac{dr^2}{1-\frac{r_0^2}{r^2}}+\biggl(1-\frac{r_0^2}{r^2}\biggr)r_0^2\, d\phi^2\,,
\end{equation}
which is indeed the same as the metric~\eqref{bonstat} of the bubble of nothing for $n=2$.

Observe that the $\phi$ circles in~\eqref{5dbubble} cannot reach arbitrarily large sizes but become a compact direction at infinity. This is a consequence of focusing on the region close to the disk at $x=1$, which limits the growth of these circles.

One might wonder whether rotating black rings, with the rotation adjusted to balance the tension and gravitational self-attraction, have a limit to the bubble of nothing. The answer is no: in the limit where the rotating ring becomes very fat, it approaches a singular, horizonless solution instead of the non-singular geometry~\eqref{5dbubble}.

\subsection{Weyl metrics and rod structures}

All other solutions in this article will be presented as vacuum Weyl metrics, using cylindrical coordinates 
\begin{equation}
\label{static-metric}
{ds}^2 = f(\rho,z) \bigl({d\rho}^2 + {dz}^2\bigr) + g_{ab}(\rho,z) {dx}^a {dx}^b \,.
\end{equation}
Readers who are familiar with this class of solutions may skip to Sec.~\ref{subsec:bbhbub}, but for those who are not, we provide a brief overview of their structure---for more complete expositions, we refer to~\cite{Belinski:2001ph,emparan-weyl,Emparan:2008eg,Harmark:2004rm}.
Their main feature is the presence of $D-2$ Killing coordinates $x^a$: in four dimensions they are $(t,\phi)$, and in five dimensions they include an additional angle, $(t,\phi,\psi)$.

For all the solutions presented here, the metric $g_{ab}(\rho,z)$ along the Killing directions will be diagonal.
Static and axisymmetric solutions can then be systematically constructed by specifying a set of rod-like sources along the $z$ axis for the three-dimensional Newtonian potentials associated to the metric functions $g_{ab}$; they are not physical rods, but coordinate singularities in the axis $\rho=0$ of the Weyl metrics.  Their importance derives from the fact that, given the rod distribution, the form of $g_{ab}(\rho,z)$ directly follows from a simple algebraic construction. 
Subsequently, $f(\rho,z)$ can be obtained by a line integral in the case of diagonal metrics, and more generally by the inverse scattering method~\cite{Belinski:2001ph}. The rods (with linear density $1/2$) are specified along each direction $x^a$, in such a way that at every value of $z$ there is a rod along one and only one of the directions. 

The rod structure provides an easy diagrammatic way to interpret static (or more generally stationary) axisymmetric solutions. 
On a rod along a direction $x^a$, the corresponding Killing vector has a fixed point set.
When we have an angular Killing vector, such as $\partial_{\phi}$ or $\partial_{\psi}$, then the corresponding circles shrink to zero size at the rod, and the periodicity of the angle must be appropriately chosen in order to avoid conical singularities. The regularity condition on $x^a\sim x^a + \Delta x^a$ at any given rod is
\begin{equation}\label{nocone}
    \Delta x^a = 2\pi \lim_{\rho\to 0}\rho \sqrt{\frac{f}{g_{aa}}}\,.
\end{equation}
When the Killing vector is timelike $\partial_t$, the rod represents a horizon, and through Euclidean continuation $t\to i\tau$, Eq.~\eqref{nocone} gives its associated temperature $T=\Delta\tau ^{-1}$.
If the rod is finite, it defines an event horizon, while infinite rods are generically associated to accelerating horizons, such as Rindler ones.

The topology of the solutions can also be inferred from the rod structure. If there is a rod along a direction $x^a$, the other directions $x^b$ are fibered along the corresponding portion of the axis. At a point where rods along $x^a$ and $x^b$ meet, the two fibers shrink to zero. As a result, the solutions have a `bubbling' structure.

To illustrate these features we will describe the simplest examples that are relevant to us here.

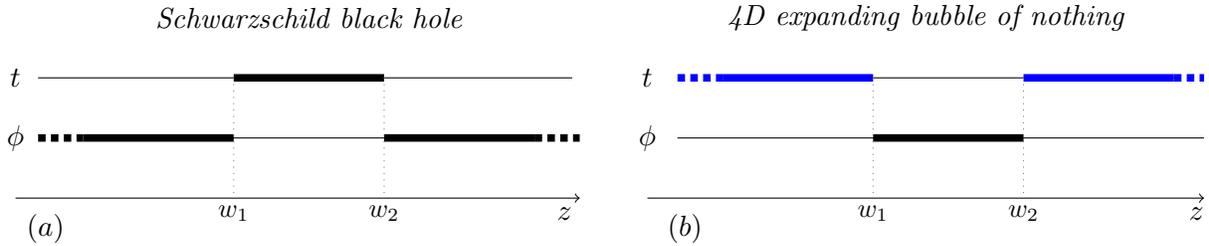
\begin{figure}[t]
\begin{tikzpicture}

\draw (-5.5,2.8) node{{{\it Schwarzschild black hole}}};
\draw (2.7,2.8) node{{{\it 4D expanding bubble of nothing}}};

\draw[black,thin] (0,2) -- (6,2);
\draw[black,thin] (-0.6,1.2) -- (6.4,1.2);
\draw[black,thin] (-9.1,2) -- (-2,2);
\draw[black,thin] (-8,1.2) -- (-2.4,1.2);

\draw[blue,dotted, line width=1mm] (-0.6,2) -- (0,2);
\draw[blue,line width=1mm] (0,2) -- (2,2);
\draw[black,line width=1mm] (2,1.2) -- (4,1.2);
\draw[blue,line width=1mm] (4,2) -- (6,2);
\draw[blue, dotted, line width=1mm] (6,2) -- (6.4,2);

\draw[black,dotted, line width=1mm] (-9.1,1.2) -- (-8.5,1.2);
\draw[black,line width=1mm] (-8.5,1.2) -- (-6.5,1.2);
\draw[black,line width=1mm] (-6.5,2) -- (-4.5,2);
\draw[black,line width=1mm] (-4.5,1.2) -- (-2.5,1.2);
\draw[black,dotted, line width=1mm] (-2.5,1.2) -- (-1.9,1.2);

\draw[gray,dotted] (2,2) -- (2,0.4);
\draw[gray,dotted] (4,2) -- (4,0.4);
\draw[gray,dotted] (-4.5,2) -- (-4.5,0.4);
\draw[gray,dotted] (-6.5,2) -- (-6.5,0.4);

\draw (-9,0) node{$(a)$};
\draw (-0.5,0) node{$(b)$};
\draw (2,0.2) node{{\small $w_1$}};
\draw (-4.5,0.2) node{{\small $w_2$}};

\draw (-6.5,0.2) node{{\small $w_1$}};
\draw (4,0.2) node{{\small $w_2$}};
\draw (6.3,0.2) node{$z$};
\draw (-2.1,0.2) node{$z$};

\draw (-1,2) node{$t$};
\draw (-1,1.2) node{$\phi$};

\draw (-9.4,2) node{$t$};
\draw (-9.4,1.2) node{$\phi$};

\draw[black,->] (-1,0.4) -- (6.4,0.4);
\draw[black,->] (-9.4,0.4) -- (-1.9,0.4);

\end{tikzpicture}
\caption{\small $(a)$ rod diagram for the Schwarzschild black hole.
The finite timelike rod defines the black hole horizon (a sphere $S^2$, from the fibration of the $\phi$ parallel circles over the segment $w_1<z<w_2$). Exchanging $t\leftrightarrow\phi$ gives $(b)$: rod diagram of the expanding bubble of nothing.
The semi-infinite timelike rods represent the bubble acceleration horizons (two of them, each with topology $\mathbb{R}^2$). Here and in the following figures, acceleration horizons are pictured in blue.}
\label{1bh-1bubble}
\end{figure}

\paragraph{4D.}
The Schwarzschild black hole rod structure is given by a finite timelike rod and two semi-infinite spacelike rods (Fig.~\ref{1bh-1bubble}(a)). The four-dimensional bubble of nothing is the double Wick-rotated version of the Schwarzschild metric, thus its rod structure is consistently given by exchanging the $t$ and $\phi$ rods of the previous solution (Fig.~\ref{1bh-1bubble}(b)). We see from their respective timelike rods that in the Schwarzschild solution the horizon is finite, while the bubble of nothing possesses two infinite acceleration horizons.

The corresponding metrics are given by
\begin{align}
\label{gschbub}
g_{ab}^\text{Schwarz}dx^a dx^b & = -\frac{\mu_1}{\mu_2}\, dt^2 + \rho^2 \frac{\mu_2}{\mu_1}\, d\phi^2 \,, \\
g_{ab}^\text{bubble}dx^a dx^b & = -\rho^2\frac{\mu_2}{\mu_1}\, dt^2+ \frac{\mu_1}{\mu_2}\, d\phi^2 \,,
\end{align}
and, in both cases,
\begin{equation}
\label{fschbub}
f = C_f \frac{ 4\mu_1\mu_2^3}{\mu_{12}W_{11}W_{22}} \,.
\end{equation}
Here, and in the following, we introduce
\begin{equation}
\label{muij-Wij}
\mu_i = w_i-z+\sqrt{\rho^2 + (z-w_i)^2} \,, \qquad 
\mu_{ij} = (\mu_{i}-\mu_{j})^2  \,, \qquad
W_{ij} = \rho^2 + \mu_i\mu_j \,.
\end{equation}
The parameters $w_i$, chosen in increasing order, specify the rod endpoints, and they define the physical properties of the metric:
the position of the horizons, the size and mass of the black holes, and the rotation axes.
The parameter $C_f$ is an arbitrary gauge constant. It corresponds to a rescaling of $\rho$ and $z$, and it can be chosen, without loss of generality, to fix the normalization of one of the Killing directions, for instance, setting the periodicity of one of the angles to any prescribed value, such as canonical periodicity $2\pi$.

It is now straightforward to verify that taking
\begin{equation}
w_1 = z_0-\frac{r_0}2 \,, \qquad
w_2 = z_0+\frac{r_0}2 \,, \qquad
C_f=r_0^2 \,,
\end{equation}
and defining
\begin{equation}
\rho = \sqrt{r(r-r_0)}\sin\theta \,, \qquad
z = z_0+\left( r-\frac{r_0}2\right)\cos\theta \,,
\end{equation}
in~\eqref{gschbub} and~\eqref{fschbub}, we recover
\begin{align}
g_{ab}^\text{Schwarz}dx^a dx^b & = -\biggl( 1-\frac{r_0}{r}\biggr) dt^2 + r^2\sin^2\theta d\phi^2 \,, \\
\label{4dbub}
g_{ab}^\text{bubble}dx^a dx^b & = -r^2\sin^2\theta dt^2+ \biggl( 1-\frac{r_0}{r}\biggr) d\phi^2 \,,
\end{align}
and
\begin{equation}
\label{f-4d}
f(\rho,z) \bigl({d\rho}^2 + {dz}^2\bigr) =
\frac{dr^2}{1-\frac{r_0}{r}}+r^2 d\theta^2 \,,
\end{equation}
which are the conventional forms of the Schwarzschild and bubble solutions (up to possible constant rescalings of the Killing coordinates $t$ and $\phi$). They are obviously equivalent under $t\leftrightarrow\phi$. The form of the bubble of nothing in \eqref{bonstat} is recovered by rescaling $\phi$ by $r_0$,\footnote{We could have achieved this by adequately choosing $C_f$, but, in general, we will not take $\phi$ to be canonically normalized with $\phi\sim \phi+2\pi$, but rather its periodicity will be suitably adjusted.} and setting $\cos\theta=\xi$.

Finally, observe that if in either of the solutions we send one of the rod endpoints, $w_1$ or $w_2$, to infinity while keeping the other fixed, then we recover the geometry of Rindler space, with an infinite acceleration horizon.
The Minkowski spacetime can be obtained when both the poles are simultaneously pushed infinitely far away in opposite directions, i.e.~$w_1 \to -\infty $ and $w_2 \to \infty$.

\begin{figure}[t]
\begin{tikzpicture}

\draw (-5.5,2.8) node{{{\it 5D black hole}}};
\draw (2.7,2.8) node{{{\it 5D expanding bubble of nothing}}};

\draw[black,thin] (0,2) -- (6.4,2);
\draw[black,thin] (-0.6,1.2) -- (6.4,1.2);
\draw[black,thin] (-0.6,0.4) -- (6,0.4);
\draw[black,thin] (-9.1,2) -- (-2,2);
\draw[black,thin] (-8,1.2) -- (-2,1.2);
\draw[black,thin] (-9.1,0.4) -- (-2.4,0.4);

\draw[blue,dotted, line width=1mm] (-0.6,2) -- (0,2);
\draw[blue,line width=1mm] (0,2) -- (2,2);
\draw[black,line width=1mm] (2,1.2) -- (4,1.2);
\draw[black,line width=1mm] (4,0.4) -- (6,0.4);
\draw[black, dotted, line width=1mm] (6,0.4) -- (6.4,0.4);

\draw[black,dotted, line width=1mm] (-9.1,1.2) -- (-8.5,1.2);
\draw[black,line width=1mm] (-8.5,1.2) -- (-6.5,1.2);
\draw[black,line width=1mm] (-6.5,2) -- (-4.5,2);
\draw[black,line width=1mm] (-4.5,0.4) -- (-2.5,0.4);
\draw[black,dotted, line width=1mm] (-2.5,0.4) -- (-1.9,0.4);

\draw[gray,dotted] (2,2) -- (2,-0.4);
\draw[gray,dotted] (4,2) -- (4,-0.4);
\draw[gray,dotted] (-4.5,2) -- (-4.5,-0.4);
\draw[gray,dotted] (-6.5,2) -- (-6.5,-0.4);

\draw (-9,-0.8) node{$(a)$};
\draw (-0.5,-0.8) node{$(b)$};
\draw (2,-0.6) node{{\small $w_1$}};
\draw (-4.5,-0.6) node{{\small $w_2$}};
\draw (-6.5,-0.6) node{{\small $w_1$}};
\draw (4,-0.6) node{{\small $w_2$}};
\draw (6.3,-0.6) node{$z$};
\draw (-2.1,-0.6) node{$z$};

\draw (-1,2) node{$t$};
\draw (-1,1.2) node{$\phi$};
\draw (-1,0.4) node{$\psi$};

\draw (-9.4,2) node{$t$};
\draw (-9.4,1.2) node{$\phi$};
\draw (-9.4,0.4) node{$\psi$};

\draw[black,->] (-1,-0.4) -- (6.4,-0.4);
\draw[black,->] (-9.4,-0.4) -- (-1.9,-0.4);

\end{tikzpicture}
\caption{\small $(a)$: rod diagram for the 5D Schwarzschild-Tangherlini black hole.
The finite timelike rod defines the black hole horizon (a sphere $S^3$, fibering $\phi$ and $\psi$ circles over $w_1<z<w_2$). Exchanging $t \leftrightarrow \phi$ gives $(b)$: rod diagram of the five-dimensional expanding bubble of nothing.
The timelike semi-infinite rod represents the bubble acceleration horizon (which is connected, with topology $S^1\times \mathbb{R}^2$: the $\psi$ circles are trivially fibered over $-\infty<z<w_1$).}
\label{fig:5D-1bh-1bubble}
\end{figure}
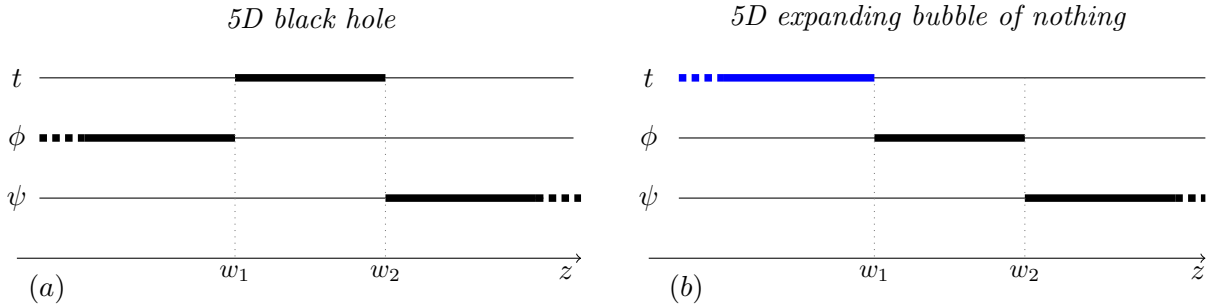

\paragraph{5D.} The previous analysis has a straightforward counterpart in five dimensions.
The rod structures of the Schwarzschild-Tangherlini black hole and the five-dimensional expanding bubble are given in Fig.~\ref{fig:5D-1bh-1bubble}, which makes evident that they are related by a double-Wick rotation that effectively exchanges $t$ and $\phi$.

These rod structures dictate that
\begin{align}
\label{g_5D_bh_bubble}
g_{ab}^\text{Tang} dx^a dx^b & =
-\frac{\mu_1}{\mu_2} {dt}^2
+ \frac{\rho^2}{\mu_1} {d\phi}^2
+ \mu_2 {d\psi}^2 \,, \\
g_{ab}^\text{5D-bubble} dx^a dx^b & =
-\frac{\rho^2}{\mu_1} {dt}^2
+ \frac{\mu_1}{\mu_2} {d\phi}^2
+ \mu_2 {d\psi}^2 \,, \label{g5dbubble}
\end{align}
while $f(\rho,z)$ is again identical for both spacetimes
\begin{equation}
f = C_f \, \frac{\mu_2 W_{12}}{W_{11}W_{22}} \,.
\end{equation}
To express the Schwarzschild-Tangherlini black hole in spherical coordinates
\begin{equation}
ds^2 = - \biggl(1-\frac{r_0^2}{r^2} \biggr)dt^2 + \frac{dr^2}{1-\frac{r_0^2}{r^2}} + r^2 d\theta^2 +  r^2 \sin^2{\theta} d\phi^2+ r^2 \cos^2{\theta} d\psi^2 \,,
\end{equation}
we choose
\begin{equation}
w_1 = z_0 - \frac{r_0^2}{4} \,, \qquad
w_2 = z_0 + \frac{r_0^2}{4} \,, \qquad
C_f =1 \,,
\end{equation}
and change
\begin{equation} \label{rho-z-tangherlini}
\rho = \frac{1}{2} r \sqrt{r^2-r_0^2} \sin{2\theta} \,, \qquad
z = z_0 + \frac{1}{4} \bigl(2r^2-r_0^2 \bigr)\cos{2\theta} \,.
\end{equation}
Similarly, the five-dimensional expanding bubble in spherical coordinates takes the form
\begin{equation}
ds^2 = -r^2 \cos^2{\theta}dt^2 + \frac{dr^2}{1-\frac{r_0^2}{r^2}} + r^2 d\theta^2 + \biggl(1-\frac{r_0^2}{r^2} \biggr) d\phi^2+ r^2 \sin^2{\theta} d\psi^2 \,.
\end{equation}
When we rescale $\phi$ by $r_0$ and set $\sin\theta=\xi$ we recover \eqref{5dbubble}.

\subsection{From binary black hole to bubble}\label{subsec:bbhbub}

\begin{figure}
\centering
\begin{tikzpicture}

\draw (-7.5,2.6) node{{{\it Two black holes}}};

\draw[black,thin] (-10.2,1.8) -- (-4.5,1.8);
\draw[black,thin] (-9,1) -- (-5,1);

\draw[black, dotted, line width=1mm] (-10.3,1) -- (-10,1);
\draw[black,line width=1mm] (-10,1) -- (-9,1);
\draw[black,line width=1mm] (-9,1.8) -- (-8,1.8);
\draw[black,line width=1mm] (-8,1) -- (-7,1);
\draw[black,line width=1mm] (-7,1.8) -- (-6,1.8);
\draw[black,line width=1mm] (-6,1) -- (-5,1);
\draw[black,dotted, line width=1mm] (-5,1) -- (-4.5,1);

\draw[gray,dotted] (-9,1.8) -- (-9,0.2);
\draw[gray,dotted] (-8,1.8) -- (-8,0.2);
\draw[gray,dotted] (-7,1.8) -- (-7,0.2);
\draw[gray,dotted] (-6,1.8) -- (-6,0.2);

\draw (-10.2,-0.1) node{{\small $(a)$}};
\draw (-9,-0) node{{\small $w_1$}};
\draw (-8,-0) node{{\small $w_2$}};
\draw (-7,-0) node{{\small $w_3$}};
\draw (-6,-0) node{{\small $w_4$}};
\draw (-4.5,-0) node{$z$};

\draw (-10.6,1.8) node{$t$};
\draw (-10.6,1) node{$\phi$};

\draw[black,->] (-10.5,0.2) -- (-4.5,0.2);


\draw[black,->] (-4,1) -- (-2.5,1)
node[midway, above, sloped] {{\small $w_1\to-\infty$}}
node[midway, below, sloped] {{\small $w_4\to\infty$}};


\draw (1.3,2.6) node{{{\it 4D Bubble of nothing}}};

\draw[black,thin] (-1,1.8) -- (3.5,1.8);
\draw[black,thin] (-1.8,1) -- (4.3,1);

\draw[blue, dotted, line width=1mm] (-1.8,1.8) -- (-1,1.8);
\draw[blue,line width=1mm] (-1,1.8) -- (0.5,1.8);
\draw[black,line width=1mm] (0.5,1) -- (2,1);
\draw[blue,line width=1mm] (2,1.8) -- (3.5,1.8);
\draw[blue,dotted, line width=1mm] (3.5,1.8) -- (4.3,1.8);

\draw[gray,dotted] (0.5,1.8) -- (0.5,0.2);
\draw[gray,dotted] (2,1.8) -- (2,0.2);

\draw (-1.8,-0.1) node{{\small $(b)$}};
\draw (0.5,-0) node{{\small $w_2$}};
\draw (2,-0) node{{\small $w_3$}};

\draw (-2.1,1.8) node{$t$};
\draw (-2.1,1) node{$\phi$};
\draw (4.5,-0) node{$z$};

\draw[black,->] (-2.1,0.2) -- (4.5,0.2);

\end{tikzpicture}
\caption{{\small $(a)$ Rod diagram for the Bach--Weyl static binary black hole configuration~\cite{bach-weyl}.
The thick timelike rods represent the two black hole horizons. By sending the rod endpoints $w_1\to-\infty$, $w_4\to\infty$, with $w_2$ and $w_3$ fixed, we recover the same diagram as in the bubble of nothing Fig.~\ref{1bh-1bubble}(b). This limit makes the two black holes infinitely large, while keeping the separation between them finite, as is illustrated in panel $(b)$.}}
\label{fig:bach-weyl}
\end{figure}
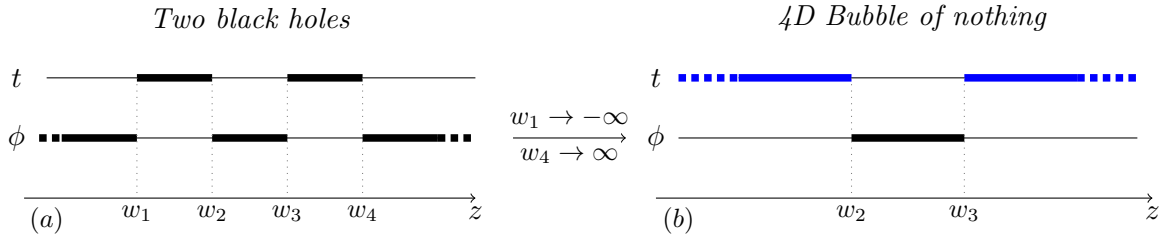

Now let us consider the Bach--Weyl solution, with 
\begin{equation}
\label{bach-weyl-metric}
g_{ab} dx^a dx^b =
- \frac{\mu_1\mu_3}{\mu_2\mu_4} d\tilde{t}^2 + \rho^2 \frac{\mu_2 \mu_4}{\mu_1 \mu_3} d\tilde{\phi}^2 \ \,, \qquad
f = \frac{16 \tilde{C}_f \ \mu_1^3 \mu_2^5 \mu_3^3 \mu_4^5}{\mu_{12} \mu_{14} \mu_{23} \mu_{34} W_{13}^2 W_{24}^2 W_{11} W_{22} W_{33} W_{44}} \,,
\end{equation}
which describes two Schwarzschild black holes aligned along the $z$-axis, and whose rod diagram is pictured in Fig.~\ref{fig:bach-weyl}(a).
By appropriately choosing $\tilde{C}_f$ to be
\begin{equation}
\tilde{C}_f = 16 (w_1-w_2)^2 (w_1-w_3)^2 (w_2-w_4)^2 (w_3-w_4)^2 \,,
\end{equation}
we make the segment $w_2<z<w_3$ of the axis in between the black holes regular, while keeping the standard periodicity of the azimuthal angle $\Delta\phi = 2\pi$. On the other hand, along the semi-infinite axes from the black holes towards $z\to\pm\infty$ there are conical deficits. These can be regarded as cosmic strings that keep the black holes apart.

The rod diagram makes manifest how this solution is connected to other black hole/bubble configurations, either via double-Wick rotations that exchange the $t$ and $\phi$ rods,\footnote{The Bach-Weyl solution is the double-Wick rotation of the single black hole in the bubble \cite{horowitz-maeda}.} or by taking limits where rod endpoints merge or are sent to infinity. For our purposes here, we observe that by simply sending the rod endpoints $w_1\to-\infty$ and $w_4\to\infty$, with $w_2$ and $w_3$ fixed, we recover the rod diagram of the 4D bubble of nothing, Fig.~\ref{1bh-1bubble}(b). When we do so, we make the two black holes infinitely large, while maintaining fixed the separation between them. This is precisely the type of limit that we discussed in the introduction (Fig.~\ref{fig:bubblebhs}). The cosmic strings collapse the space along the outer axes creating a conical deficit angle of $2\pi$, but this is not a problem since this part of the geometry is pushed away to infinity.

To see that the limit works correctly, not only with the rods but also in the entire metric, we conveniently place the bubble poles symmetrically at $w_1=-z_b$ and $w_4=z_b$.
Then, we rescale 
\begin{equation}
\tilde{t} = (2z_b) t \,, \qquad
\tilde{\phi} = \frac{\phi}{2z_b} \,,
\end{equation}
so that the metric \eqref{bach-weyl-metric} remains finite when we send $z_b \to \infty$. 
One can readily verify that, after rescaling $\tilde{C}_f=C_f/4$ to take into account that in \eqref{gschbub} $\phi$ has periodicity $4\pi$, the bubble of nothing in the form of~\eqref{gschbub} and~\eqref{fschbub} is recovered.

We have then proven that the gravitational field of the expanding bubble is indeed the same as that between two very large black holes.

\subsection{From black ring to bubble, Weyl style}\label{subsec:ringtobubble2}

It is now easy to see how the rod diagrams also make transparent the limit from the static black ring to the expanding bubble of nothing, which we discussed using other coordinates in Sec.~\ref{subsec:ringtobubble}. 

\begin{figure}[t]
\centering
\begin{tikzpicture}

\draw (-5.7,2.6) node{{{\it Black ring}}};
\draw (3.2,2.6) node{{{\it 5D Bubble of nothing}}};

\draw[black,thin] (-9.1,2) -- (-2,2);
\draw[black,thin] (-8,1.2) -- (-2,1.2);
\draw[black,thin] (-9.1,0.4) -- (-2.4,0.4);

\draw[black,thin] (1,2) -- (6.1,2);
\draw[black,thin] (0.4,1.2) -- (6.1,1.2);
\draw[black,thin] (0.4,0.4) -- (5.5,0.4);

\draw[black,dotted, line width=1mm] (-9.1,1.2) -- (-8.5,1.2);
\draw[black,line width=1mm] (-8.5,1.2) -- (-7,1.2);
\draw[black,line width=1mm] (-7,2) -- (-5.3,2);
\draw[black,line width=1mm] (-5.3,1.2) -- (-4.2,1.2);
\draw[black,line width=1mm] (-4.2,0.4) -- (-2.5,0.4);
\draw[black,dotted, line width=1mm] (-2.5,0.4) -- (-1.9,0.4);

\draw[blue,dotted, line width=1mm] (0.4,2) -- (1,2);
\draw[blue,line width=1mm] (1,2) -- (2.7,2);
\draw[black,line width=1mm] (2.7,1.2) -- (3.8,1.2);
\draw[black,line width=1mm] (3.8,0.4) -- (5.5,0.4);
\draw[black, dotted, line width=1mm] (5.5,0.4) -- (6.1,0.4);

\draw[gray,dotted] (-7,2) -- (-7,-0.4);
\draw[gray,dotted] (-5.3,2) -- (-5.3,-0.4);
\draw[gray,dotted] (-4.2,2) -- (-4.2,-0.4);

\draw[gray,dotted] (2.7,2) -- (2.7,-0.4);
\draw[gray,dotted] (3.8,2) -- (3.8,-0.4);

\draw (-9,-0.8) node{$(a)$};
\draw (0.5,-0.8) node{$(b)$};

\draw (-7,-0.6) node{{\small $w_0$}};
\draw (-5.3,-0.6) node{{\small $w_1$}};
\draw (-4.2,-0.6) node{{\small $w_2$}};

\draw (2.7,-0.6) node{{\small $w_1$}};
\draw (3.8,-0.6) node{{\small $w_2$}};

\draw (6.3,-0.6) node{$z$};
\draw (-2.1,-0.6) node{$z$};

\draw (-9.4,2) node{$t$};
\draw (-9.4,1.2) node{$\phi$};
\draw (-9.4,0.4) node{$\psi$};

\draw (0.1,2) node{$t$};
\draw (0.1,1.2) node{$\phi$};
\draw (0.1,0.4) node{$\psi$};

\draw[black,->] (-9.4,-0.4) -- (-1.9,-0.4);
\draw[black,->] (0.1,-0.4) -- (6.4,-0.4);

\draw[black,->] (-1.6,0.8) -- (-0.4,0.8) node[midway, below, sloped] {{\small $w_0\to-\infty$}};

\end{tikzpicture}
\caption{{\small $(a)$ Rod diagram for the static black ring.
The thick timelike rod represents the black ring horizon, with topology $S^1\times S^2$. $(b)$ By sending the rod endpoint $w_0\to-\infty$, with $w_1$ and $w_2$ fixed, we recover the bubble of nothing in Fig.~\ref{fig:5D-1bh-1bubble}(b). This limit makes the black ring very fat, while keeping its hole finite, as was illustrated in Fig.~\ref{fig:bubblebhs}.}}
\label{fig:statring}
\end{figure}
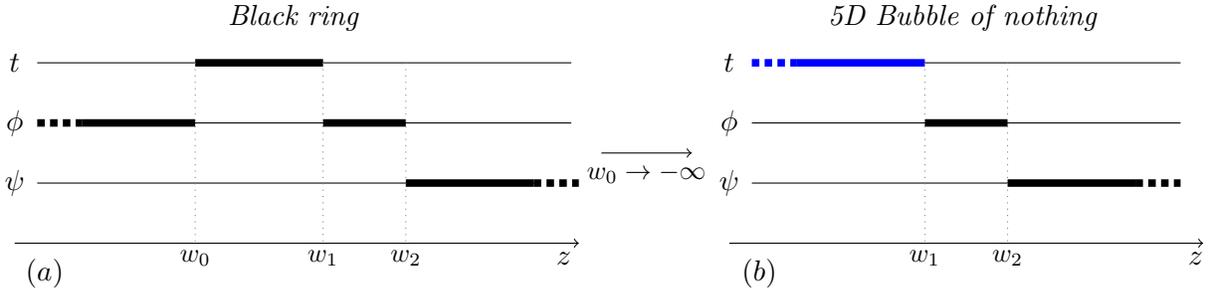

Fig.~\ref{fig:statring}(a) shows the rod diagram for the static black ring.
The Weyl form of the metric that follows from the diagram is
\begin{equation}
g_{ab} dx^a dx^b =
-\frac{\mu_0}{\mu_1} {d\tilde{t}}^2
+ \rho^2 \frac{\mu_1}{\mu_0\mu_2} {d\tilde{\phi}}^2
+ \mu_2 {d\psi}^2 \,, \qquad
f = C_f
\frac{\mu_2 W_{01}^2 W_{12}}{W_{02} W_{00} W_{11} W_{22}} \,.
\end{equation}
The horizon of the black ring, with topology $S^1\times S^2$, lies at $w_0<z<w_1$, while the `hole' of the ring is in the region $w_1<z<w_2$.
If we send $w_0\to -\infty$  keeping all other rod endpoints fixed---hence making the ring very fat while its hole remains finite---we recover the same diagram as for the expanding bubble of nothing in Fig.~\ref{fig:5D-1bh-1bubble}(b). In the metric, this requires a suitable rescaling of $t$ and $\phi$, similarly to what happens in the 4D case.
The required rescalings are
\begin{equation}
\tilde{t} = \sqrt{2|w_0|} t \,, \qquad
\tilde{\phi} = \frac{\phi}{\sqrt{2|w_0|}} \,.
\end{equation}

\subsection{5D black hole binaries and bubbles}\label{subsec:5dbbh}

We shall briefly mention how a limit can be taken in a five-dimensional black hole binary that is asymptotically flat (save for possible conical defect membranes) to yield a different kind of five-dimensional expanding bubble.

\begin{figure}[t]
\centering
\begin{tikzpicture}

\draw (-5.7,2.6) node{{{\it 5D black hole binary}}};
\draw (3.2,2.6) node{{{\it Double bubble of nothing}}};

\draw[black,thin] (-9.1,2) -- (-2,2);
\draw[black,thin] (-9.1,1.2) -- (-2.6,1.2);
\draw[black,thin] (-8.7,0.4) -- (-2,0.4);

\draw[black,thin] (1,2) -- (5.5,2);
\draw[black,thin] (0.4,1.2) -- (6.1,1.2);
\draw[black,thin] (0.4,0.4) -- (6.1,0.4);

\draw[black,dotted, line width=1mm] (-9.1,0.4) -- (-8.7,0.4);
\draw[black,line width=1mm] (-8.7,0.4) -- (-8.1,.4);
\draw[black,line width=1mm] (-8.1,2) -- (-7,2);
\draw[black,line width=1mm] (-7,1.2) -- (-6,1.2);
\draw[black,line width=1mm] (-6,0.4) -- (-4.8,0.4);
\draw[black,line width=1mm] (-4.8,2) -- (-3.5,2);
\draw[black,line width=1mm] (-3.5,1.2) -- (-2.6,1.2);
\draw[black, dotted, line width=1mm] (-2.6,1.2) -- (-2,1.2);

\draw[blue,dotted, line width=1mm] (0.4,2) -- (.9,2);
\draw[blue,line width=1mm] (.9,2) -- (2.1,2);
\draw[black,line width=1mm] (2.1,1.2) -- (3.1,1.2);
\draw[black,line width=1mm] (3.1,0.4) -- (4.3,0.4);
\draw[blue,line width=1mm] (4.3,2) -- (5.5,2);
\draw[blue, dotted, line width=1mm] (5.5,2) -- (6.1,2);

\draw[gray,dotted] (-8.1,2) -- (-8.1,-0.4);
\draw[gray,dotted] (-7,2) -- (-7,-0.4);
\draw[gray,dotted] (-6,2) -- (-6,-0.4);
\draw[gray,dotted] (-4.8,2) -- (-4.8,-0.4);
\draw[gray,dotted] (-3.5,2) -- (-3.5,-0.4);

\draw[gray,dotted] (2.1,2) -- (2.1,-0.4);
\draw[gray,dotted] (3.1,2) -- (3.1,-0.4);
\draw[gray,dotted] (4.3,2) -- (4.3,-0.4);

\draw (-9,-0.8) node{$(a)$};
\draw (0.6,-0.8) node{$(b)$};

\draw (-8.1,-0.6) node{{\small $w_0$}};
\draw (-7,-0.6) node{{\small $w_1$}};
\draw (-6,-0.6) node{{\small $w_2$}};
\draw (-4.8,-0.6) node{{\small $w_3$}};
\draw (-3.5,-0.6) node{{\small $w_4$}};

\draw (2.1,-0.6) node{{\small $w_1$}};
\draw (3.1,-0.6) node{{\small $w_2$}};
\draw (4.3,-0.6) node{{\small $w_3$}};

\draw (6.3,-0.6) node{$z$};
\draw (-2.1,-0.6) node{$z$};

\draw (-9.4,2) node{$t$};
\draw (-9.4,1.2) node{$\phi$};
\draw (-9.4,0.4) node{$\psi$};

\draw (0.1,2) node{$t$};
\draw (0.1,1.2) node{$\phi$};
\draw (0.1,0.4) node{$\psi$};

\draw[black,->] (-9.4,-0.4) -- (-1.9,-0.4);
\draw[black,->] (0.1,-0.4) -- (6.4,-0.4);

\draw[black,->] (-1.6,0.8) -- (-0.4,0.8) node[midway, above, sloped] {{\small $w_0\to-\infty$}};
\draw[black,->] (-1.6,0.8) -- (-0.4,0.8) node[midway, below, sloped] {{\small $w_4\to \infty$}};

\end{tikzpicture}
\caption{{\small $(a)$ Rod diagram for the five-dimensional black hole binary of \cite{Tan:2003jz}. $(b)$ Limit to the expanding bubble of nothing of Sec.~4.7 in \cite{emparan-weyl}, which is asymptotic to a space compactified on a two-torus. The space in between the horizons, $w_1<z<w_3$, consists of two topological disks $D_2$, orthogonal to each other and touching at their origins (at $z=w_2$). In the maximal analytic extension, these become two orthogonal $S^2$ that touch at their poles.}}
\label{fig:5dbbh}
\end{figure}
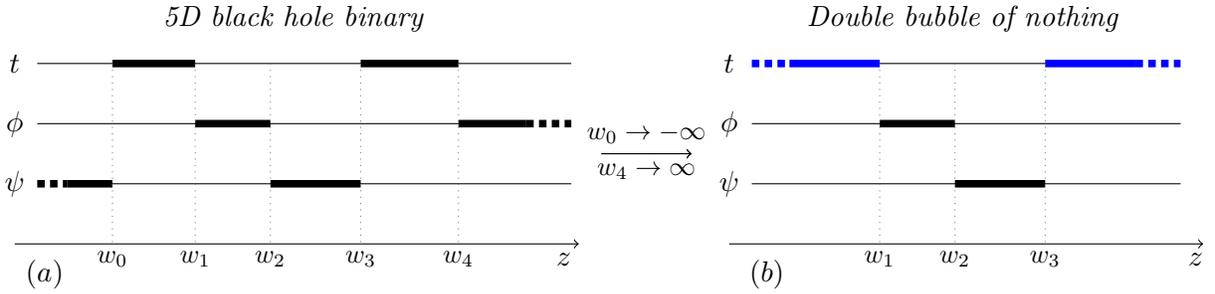

The Weyl formalism allows to combine two five-dimensional Schwarzschild-Tangherlini black holes with the rod structure in Fig.~\ref{fig:5dbbh}(a). This system was studied in \cite{Tan:2003jz}. Since the two black holes lie along different axes, they cannot be regarded as collinear. Nevertheless, the solution is asymptotically flat, as follows from the presence of one semi-infinite rod along $\phi$ and another along $\psi$. Now take, as in the previous examples, the limit where the two black holes become infinitely large, making their timelike rods semi-infinite. The result is the system in Fig.~\ref{fig:5dbbh}(b).

This geometry was analyzed in \cite{emparan-weyl}, where one can find the explicit solution (see Sec.~4.7 there). Here we shall only describe its main properties. Conical singularities can be avoided along all the axes, and the solution is identified as an expanding bubble of nothing. In contrast to the simpler five-dimensional bubble of \eqref{5dbubble} (and \eqref{g5dbubble}), where the minimal cycle (the bubble) is a sphere $S^2$, in this case it is made of two orthogonal $S^2$, i.e., the meridian lines of one sphere are orthogonal to the meridian lines of the other, and the parallel lines of one lie along $\phi$ and of the other along $\psi$. The two spheres touch each other at both their north and south poles. Furthermore, the single bubble in \eqref{5dbubble} asymptotically has one compact circle, while the double bubble in Fig.~\ref{fig:5dbbh}(b) has two, and therefore represents a Kaluza-Klein compactification from five to three dimensions. Each of the two $S^2$ is responsible for the compactification of one of the two circle directions. The solution also differs from the four-dimensional bubble \eqref{4dbub}, in that the two acceleration horizons here are not symmetric: the $\phi$ and $\psi$ circles close off at one or the other horizon, and their accelerations can be different. Indeed, the two $S^2$ can have different sizes.

The two five-dimensional black holes can also be combined in a different fashion with Kaluza-Klein asymptotics \cite{emparan-weyl,Elvang:2002br}. The configuration has a limit to a `bubble string', i.e., the direct product of the 4D expanding bubble and a circle. Ref.~\cite{emparan-weyl} showed that the Weyl formalism allows to generalize all of these solutions to other expanding bubbles in higher dimensions, which compactify spacetime down to three or four dimensions.

Finally, we could envisage starting from a collinear pair of 5D black holes which lie along a line that is a fixed point of $SO(3)$ rotations (and not $SO(2)$, as above). In this case, the limit of small separation would result, like in 4D, in a topologically circular $S^1$ bubble. However, these configurations (and their higher-dimensional counterparts) do not fall within the Weyl class, and they are not known in exact form.

\section{Static black hole binaries and black rings in expanding bubbles}
\label{sec:bhsfrombubbles}

In this section, we will explore some configurations in 4D and 5D that can be regularised by the presence of an expanding bubble of nothing.
First, we will consider a 4D static black hole binary system (a subcase of the Israel-Khan solution~\cite{israel-khan}). As is well known, the Bach-Weyl binary in \eqref{bach-weyl-metric} necessarily contains conical singularities on the axis $\rho=0$, either in the segment in between the two black holes, or (as we chose above) in the semi-axes towards infinity---these are, respectively, struts or strings that balance the attraction between the black holes.
We will prove how, by placing the binary within the bubble, we can remove all these singularities and thus obtain a completely regular system on and outside the event horizons.

An analogous construction is possible for the 5D static black ring. In the manner we presented this solution in Secs.~\ref{subsec:ringtobubble} and \ref{subsec:ringtobubble2}, the geometry is singular because the tension and self-attraction of the ring, which would drive it to collapse, need to be balanced by a conical-defect membrane.
Again, immersing the ring in an expanding bubble of nothing allows to balance the forces and remove all the conical singularities.

In the following we present the metrics for these systems and prove that it is possible to achieve equilibrium configurations. A more complete analysis of the physical magnitudes and of the first law of thermodynamics for black hole systems in expanding bubbles will be the subject of future work~\cite{AVinprogress}.

\subsection{4D black hole binary in equilibrium inside the expanding bubble}
\label{sec:binary+bubble}

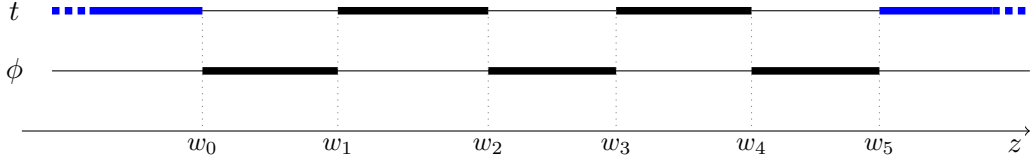
\begin{figure}[t]
\centering
\begin{tikzpicture}

\draw[black,thin] (-8.5,2) -- (2.8,2);
\draw[black,thin] (-9.5,1.2) -- (3.5,1.2);

\draw[blue, dotted, line width=1mm] (-9.5,2) -- (-9,2);
\draw[blue,line width=1mm] (-9.0,2) -- (-7.5,2);
\draw[black,line width=1mm] (-7.5,1.2) -- (-5.7,1.2);
\draw[black,line width=1mm] (-5.7,2) -- (-3.7,2);
\draw[black,line width=1mm] (-3.7,1.2) -- (-2,1.2);
\draw[black,line width=1mm] (-2,2) -- (-0.2,2);
\draw[black,line width=1mm] (-0.2,1.2) -- (1.5,1.2);
\draw[blue,line width=1mm] (1.5,2) -- (3,2);
\draw[blue, dotted, line width=1mm] (3,2) -- (3.5,2);

\draw[gray,dotted] (-7.5,2) -- (-7.5,0.4);
\draw[gray,dotted] (-5.7,2) -- (-5.7,0.4);
\draw[gray,dotted] (-3.7,2) -- (-3.7,0.4);
\draw[gray,dotted] (-2,2) -- (-2,0.4);
\draw[gray,dotted] (-0.2,2) -- (-0.2,0.4);
\draw[gray,dotted] (1.5,2) -- (1.5,0.4);

\draw (-7.5,0.2) node{{\small $w_0$}};
\draw (-5.7,0.2) node{{\small $w_1$}};
\draw (-3.7,0.2) node{{\small $w_2$}};
\draw (-2,0.2) node{{\small $w_3$}};
\draw (-0.2,0.2) node{{\small $w_4$}};
\draw (1.5,0.2) node{{\small $w_5$}};
\draw (3.3,0.2) node{$z$};

\draw (-10,2) node{$t$};
\draw (-10,1.2) node{$\phi$};

\draw[black,->] (-9.9,0.4) -- (3.5,0.4);

\end{tikzpicture}
\caption{{\small Rod diagram for a binary black hole system inside a bubble of nothing.
The black timelike rods (thick lines of the $t$ coordinate) represent the black hole horizons, while the blue timelike semi-infinite rods correspond to the bubble horizon.
This diagram corresponds to the double-Wick rotation of the three-source Israel-Khan solution~\cite{israel-khan}.}}
\label{fig:binary-bubble}
\end{figure}

Superposing the rods of the 4D bubble of nothing and the Bach-Weyl binary, we get the diagram of Fig.~\ref{fig:binary-bubble}.

The solution can be written explicitly in Weyl coordinates~\eqref{static-metric} with
\begin{subequations}
\label{binary-full}
\begin{gather}
\label{gph-binary}
g_{ab} dx^a dx^b =
-\rho^2 \frac{\mu_1 \mu_3 \mu_5}{\mu_0 \mu_2 \mu_4} {dt}^2
+ \frac{\mu_0 \mu_2 \mu_4}{\mu_1 \mu_3 \mu_5} {d\phi}^2 \,, \\
\label{fph-binary}
f = \frac{16 C_f \, \mu_0^5 \mu_1^7\mu_2^5\mu_3^7\mu_4^5\mu_5^7}{\mu_{01} \mu_{03} \mu_{05} \mu_{12} \mu_{14} \mu_{23} \mu_{25} \mu_{34} \mu_{45} W_{02}^2 W_{04}^2 W_{13}^2 W_{15}^2 W_{24}^2 W_{35}^2 W_{00} W_{11} W_{22} W_{33} W_{44} W_{55}} \,.
\end{gather}
\end{subequations}
In the limit in which the bubble horizon is pushed to infinity, for $w_0 \to -\infty$ and $w_5\to \infty$, we recover the standard Bach-Weyl binary~\eqref{bach-weyl-metric}. On the other hand the limit to the bubble can be obtained in different ways: by focusing on the bubbles in between black holes as we have done above, e.g..~taking $w_1 \to -\infty$ and $w_4 \to \infty$, or alternatively by eliminating the black holes by collapsing their rods, thus $w_1=w_2=w_3=w_4$.

In general, the geometry contains conical singularities on the $z$-axis in the intervals $(w_0,w_1)$, $(w_2,w_3)$, and  $(w_4,w_5)$, which we eliminate by imposing \eqref{nocone} on each interval. 
As we mentioned, we can choose $C_f$ (i.e., a rescaling of $f$) to set $\Delta\phi=2\pi$ without loss of generality.
Then, requiring \eqref{nocone} on $z\in(w_0,w_1)$ fixes
\begin{equation}
C_f = 2^{12}(w_0-w_2)^2 (w_1-w_2)^2 (w_2-w_3)^2 (w_0-w_4)^2(w_1-w_4)^2(w_3-w_4)^2(w_2-w_5)^2(w_4-w_5)^2 \,,
\end{equation}
while for $z\in(w_2,w_3)$ and $z\in(w_4,w_5)$ we get
\begin{subequations}\label{equilbinarybubble}
\begin{align}
\frac{(w_0-w_2)(w_2-w_3)(w_1-w_4)(w_2-w_5)}{(w_0-w_1)(w_1-w_3)(w_2-w_4)(w_1-w_5)} &= 1  \,, \\
\frac{(w_0-w_2)(w_0-w_4)(w_2-w_5)(w_4-w_5)}{(w_0-w_1)(w_0-w_3)(w_1-w_5)(w_3-w_5)} &= 1 \,.
\end{align}
\end{subequations}
These can be solved in terms of the bubble parameters $w_0$ and $w_5$, thus leaving the binary parameters $w_{1,2,3,4}$ unconstrained.
To this end, we first choose a convenient parametrization of the rod endpoints in terms of the Komar masses $M_1$, $M_2$ of the two black holes (these are half the coordinate length of the horizon rod), the coordinate distance between them, $d$, and their coordinate distances to the left and right bubble horizons, $\ell_1$ and $\ell_2$, so that
\begin{align}\label{binbubparam}
    &w_0=-\ell_1,\qquad w_1=0,\qquad w_2=2M_1,\qquad w_3=2M_1+d,\qquad \nn\\
    &w_4=2M_1+2M_2+d,\qquad w_5=2M_1+2M_2+d+\ell_2\,.
\end{align}
We then solve the equilibrium conditions \eqref{equilbinarybubble} for $\ell_1$ and $\ell_2$, to find
\begin{equation}\label{solell}
    \ell_i=\frac{\sqrt{A_i+B_i^2}-B_i}{2M_1 M_2+d(M_1+M_2)}\,,
\end{equation}
where we have defined
\begin{align}
    A_1&= d (d + 2M_1) (d + M_2) (d + 2(M_1 + M_2)) (2M_1 M_2 + d (M_1 + M_2))\,,\\
    B_1&= d^2 M_2 + 2 M_1 M_2 (M_1 + M_2) + d (M_1^2 + 3 M_1 M_2 + M_2^2)\,,
\end{align}
and $A_2$, $B_2$ are obtained by changing $1\leftrightarrow 2$. Since $\ell_1$ and $\ell_2$ in \eqref{solell} are manifestly positive when $M_1$, $M_2$, $d$ are positive, we have proven that there always exists a unique bubble, with suitably chosen position and size, that provides the necessary expansion to balance an arbitrary binary in static equilibrium (even if unstable).

It is interesting to observe that when the two black holes are very close, $d\ll M_1,M_2$, the bubble distance to them becomes
\begin{equation}
    \ell_1,\ell_2=d+O(d^2)\,,
\end{equation}
i.e., as expected, the bubble snugly hugs the binary. When the black holes are instead far apart, $d\gg M_1,M_2$, we have
\begin{equation}\label{farbin}
    \ell_1,\ell_2 =\frac{d^{3/2}}{\sqrt{M_1+M_2}}\left( 1+ O(d^{-1/2})\right)\,,
\end{equation}
which we can easily understand. The Newtonian gravitational potential between the black holes is
\begin{equation}
    V_{g}\simeq -\frac{M_1+M_2}{d}\,,
\end{equation}
and the gravitational potential from the de Sitter-like expanding space between them is (for $\ell_{1,2}\simeq \ell$)
\begin{equation}
    V_{exp}\simeq -\frac{d^2}{2\ell^2}\,,
\end{equation}
since $1/\ell^2$ acts like a cosmological constant.\footnote{The two potentials can be read from $g_{tt}$ in the weak field regime.} Then \eqref{farbin} follows from the equilibrium condition 
\begin{equation}
    \frac{\partial(V_{g}+V_{exp})}{\partial d}=0\,. 
\end{equation}

\subsection{Black ring in equilibrium inside the expanding bubble}
\label{sec:ring+bubble}

Now we insert a static black ring inside a five-dimensional bubble of nothing. Instead of the $(x,y)$ coordinates used in \eqref{bring}, we will employ Weyl coordinates. For the black ring, the explicit transformation can be found in~\cite{emparan-weyl}.

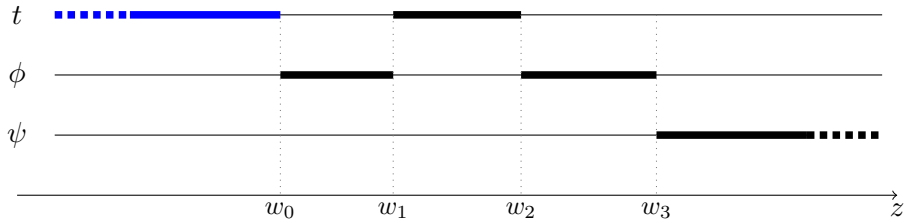
\begin{figure}[t]
\centering
\begin{tikzpicture}

\draw (-4.,2.8) node{{{\it 5D black ring inside an expanding bubble}}};

\draw[black,thin] (-9,2) -- (1,2);
\draw[black,thin] (-10,1.2) -- (1,1.2);
\draw[black,thin] (-10,0.4) -- (0,0.4);

\draw[blue, dotted, line width=1mm] (-10,2) -- (-9,2);
\draw[blue,line width=1mm] (-9,2) -- (-7,2);
\draw[black,line width=1mm] (-7,1.2) -- (-5.5,1.2);
\draw[black,line width=1mm] (-5.5,2) -- (-3.8,2);
\draw[black,line width=1mm] (-3.8,1.2) -- (-2,1.2);
\draw[black,line width=1mm] (-2,0.4) -- (0,0.4);
\draw[black, dotted,line width=1mm] (0,0.4) -- (1,0.4);

\draw[gray,dotted] (-7,2) -- (-7,-0.4);
\draw[gray,dotted] (-5.5,2) -- (-5.5,-0.4);
\draw[gray,dotted] (-3.8,2) -- (-3.8,-0.4);
\draw[gray,dotted] (-2,2) -- (-2,-0.4);

\draw (-7,-0.6) node{{\small $w_0$}};
\draw (-5.5,-0.6) node{{\small $w_1$}};
\draw (-3.8,-0.6) node{{\small $w_2$}};
\draw (-2,-0.6) node{{\small $w_3$}};
\draw (1.2,-0.6) node{$z$};

\draw (-10.5,2) node{$t$};
\draw (-10.5,1.2) node{$\phi$};
\draw (-10.5,0.4) node{$\psi$};

\draw[black,->] (-10.5,-0.4) -- (1.2,-0.4);

\end{tikzpicture}
\caption{{\small Rod diagram for the black ring inside the bubble of nothing.
A semi-infinite timelike rod for $z<w_0$ has been added to the ordinary black ring diagram.}}
\label{fig:ring-bubble}
\end{figure}

The rod diagram for the black ring is represented by the black lines in Fig.~\ref{fig:ring-bubble}, and we add the bubble by putting an extra pole and the blue line representing the bubble horizon. Incidentally, this diagram is the double Wick-rotated version of the static black Saturn~\cite{Elvang:2007rd} (see also Fig.~\ref{fig:4diagrams}).

The metric corresponding to Fig.~\ref{fig:ring-bubble} is
\begin{subequations}
\begin{align}
g_{ab} dx^a dx^b &=
-\rho^2 \frac{\mu_1}{\mu_0\mu_2} {dt}^2
+ \frac{\mu_0\mu_2}{\mu_1\mu_3} {d\phi}^2
+ \mu_3 {d\psi}^2 \,, \\
f &= C_f
\frac{\mu_3 W_{01}^2 W_{03} W_{12}^2 W_{23}}{W_{02}^2 W_{13} W_{00} W_{11} W_{22} W_{33}} \,.
\end{align}
\end{subequations}
We must eliminate conical singularities by tuning the parameters of the solution to satisfy \eqref{nocone} at every spacelike rod.
If we choose $\Delta\psi=2\pi$, we find that \eqref{nocone} is satisfied along the segment $z\in(w_3,\infty)$ by setting $C_f=1$. Next, imposing \eqref{nocone} on the $\phi$ direction along $z\in(w_0,w_1)$ and along $z\in(w_2,w_3)$, we obtain
\begin{subequations}
\label{ring-constraint}
\begin{align}
\frac{(w_1-w_0)^2(w_3-w_0)}{(w_2-w_0)^2} & = \frac{1}{2}\left(\frac{\Delta\phi}{2\pi}\right)^2 \,, \\
\frac{(w_3-w_0)(w_3-w_2)}{w_3-w_1} & = \frac{1}{2}\left(\frac{\Delta\phi}{2\pi}\right)^2 \,.
\end{align}
\end{subequations}
In order to solve these equations, we parametrize the rod endpoints as
\begin{equation}
w_0=-\ell\,,\qquad w_1 = 0 \,, \qquad
w_2 = 2\mu R^2 \,, \qquad
w_3 = (1+\mu)R^2 \,.
\end{equation}
Here $\ell$ characterizes the bubble size, while $\mu\in [0,1)$ and $R$ are the same parameters for the shape and radius of the ring as in \eqref{bring}. Eqs.~\eqref{ring-constraint} are solved with
\begin{equation}
    \ell=R^2\left(1-\mu+\sqrt{1-\mu^2}\right)\,,
\end{equation}
and\footnote{We could absorb a scale $\propto R$ in the definition of $\phi$ to make it dimensionless, as we have done before.}
\begin{equation}
    \left(\frac{\Delta\phi}{2\pi}\right)^2=2R^2\left(2+\sqrt{1-\mu^2}\right)\frac{1-\mu}{1+\mu}\,.
\end{equation}
Thus, we can always choose, in a unique way, the bubble size $\ell$ so as to balance into equilibrium an arbitrary static black ring.

\medskip

To finish this section, we shall mention that, with a straightforward exercise in rodology, which we leave to the reader, one can insert the five-dimensional binary of Fig.~\ref{fig:5dbbh}(a) inside the bubble of Fig.~\ref{fig:5dbbh}(b), and then obtain the corresponding solution (which is a limit of the ones in \cite{Tan:2003jz}). Given our previous analyses, it is natural to expect, and consistent with parameter counting, that the bubble parameters can be adjusted to balance an arbitrary binary of this kind.

\section{Other configurations}\label{sec:otherconfig}

We can extend the discussion of the previous sections to more general configurations, and play with the rods to move from one solution to another.
There are plenty of examples that can be considered, both in four and five dimensions, and even in higher dimensions \cite{emparan-weyl}.
We will consider some of them, just to give a taste of the many possibilities that are offered by the rod diagram machinery. The limits presented below on the rods diagrams work faithfully on the corresponding metrics.

\begin{figure}[t]
\centering
\begin{tikzpicture}

\draw (-6.7,2.6) node{{{\it Three black holes}}};

\draw[black,thin] (-10.2,1.8) -- (-3,1.8);
\draw[black,thin] (-9,1) -- (-4,1);

\draw[black, dotted, line width=1mm] (-10.3,1) -- (-10,1);
\draw[black,line width=1mm] (-9.9,1) -- (-9,1);
\draw[black,line width=1mm] (-9,1.8) -- (-8,1.8);
\draw[black,line width=1mm] (-8,1) -- (-7,1);
\draw[black,line width=1mm] (-7,1.8) -- (-6,1.8);
\draw[black,line width=1mm] (-6,1) -- (-5,1);
\draw[black,line width=1mm] (-5,1.8) -- (-4,1.8);
\draw[black,line width=1mm] (-4,1) -- (-3.1,1);
\draw[black,dotted, line width=1mm] (-3,1) -- (-2.8,1);

\draw[gray,dotted] (-9,1.8) -- (-9,0.2);
\draw[gray,dotted] (-8,1.8) -- (-8,0.2);
\draw[gray,dotted] (-7,1.8) -- (-7,0.2);
\draw[gray,dotted] (-6,1.8) -- (-6,0.2);
\draw[gray,dotted] (-5,1.8) -- (-5,0.2);
\draw[gray,dotted] (-4,1.8) -- (-4,0.2);

\draw (-10,-0.1) node{{\small $(a)$}};
\draw (-9,-0) node{{\small $w_0$}};
\draw (-8,-0) node{{\small $w_1$}};
\draw (-7,-0) node{{\small $w_2$}};
\draw (-6,-0) node{{\small $w_3$}};
\draw (-5,-0) node{{\small $w_4$}};
\draw (-4,-0) node{{\small $w_5$}};
\draw (-3.2,-0) node{$z$};

\draw (-10.6,1.8) node{$t$};
\draw (-10.6,1) node{$\phi$};

\draw[black,->] (-10.5,0.2) -- (-3,0.2);


\draw[black,->] (-2.6,1) -- (-1.7,1) node[midway, above, sloped] {{\scriptsize $w_0\to-\infty$}};
\draw[black,->] (-2.6,1) -- (-1.7,1) node[midway, below, sloped] {{\scriptsize $w_5\to\infty$}};


\draw (1.8,2.6) node{{{\it Single black hole inside a bubble}}};

\draw[black,thin] (-0.5,1.8) -- (4.5,1.8);
\draw[black,thin] (-1,1) -- (5,1);

\draw[blue, dotted, line width=1mm] (-1,1.8) -- (-0.5,1.8);
\draw[blue,line width=1mm] (-0.5,1.8) -- (0.5,1.8);
\draw[black,line width=1mm] (0.5,1) -- (1.5,1);
\draw[black,line width=1mm] (1.5,1.8) -- (2.5,1.8);
\draw[black,line width=1mm] (2.5,1) -- (3.5,1);
\draw[blue,line width=1mm] (3.5,1.8) -- (4.5,1.8);
\draw[blue,dotted, line width=1mm] (4.5,1.8) -- (5,1.8);

\draw[gray,dotted] (0.5,1.8) -- (0.5,0.2);
\draw[gray,dotted] (1.5,1.8) -- (1.5,0.2);
\draw[gray,dotted] (2.5,1.8) -- (2.5,0.2);
\draw[gray,dotted] (3.5,1.8) -- (3.5,0.2);

\draw (-0.7,-0.1) node{{\small $(b)$}};
\draw (0.5,-0) node{{\small $w_1$}};
\draw (1.5,-0) node{{\small $w_2$}};
\draw (2.5,-0) node{{\small $w_3$}};
\draw (3.5,-0) node{{\small $w_4$}};

\draw (-1.3,1.8) node{$t$};
\draw (-1.3,1) node{$\phi$};
\draw (4.8,-0) node{$z$};

\draw[black,->] (-1.3,0.2) -- (5,0.2);


\draw[black,->] (-6.5,-0.5) -- (-6.5,-1.5);
\draw (-5.6,-1) node{{\small $w_5\to \infty$}};

\draw[black,thin] (-10.3,-2) -- (-4,-2);
\draw[black,thin] (-9,-2.8) -- (-3.4,-2.8);
\draw[black,->] (-10.5,-3.6) -- (-3.4,-3.6);

\draw (-10.6,-2) node{$t$};
\draw (-10.6,-2.8) node{$\phi$};
\draw (-3.6,-3.8) node{$z$};
\draw (-10,-3.9) node{{\small $(c)$}};
\draw (-9,-3.8) node{{\small $w_0$}};
\draw (-8,-3.8) node{{\small $w_1$}};
\draw (-7,-3.8) node{{\small $w_2$}};
\draw (-6,-3.8) node{{\small $w_3$}};
\draw (-5,-3.8) node{{\small $w_4$}};
\draw (-6.8,-4.5) node{{ {\it Two accelerating black holes}}};

\draw[black,->] (-3.1,-1.7) -- (-1.5,-0.2) node[midway, below, sloped] {{\small $w_0\to -\infty$}};

\draw[gray,dotted] (-9,-2) -- (-9,-3.6);
\draw[gray,dotted] (-8,-2) -- (-8,-3.6);
\draw[gray,dotted] (-7,-2) -- (-7,-3.6);
\draw[gray,dotted] (-6,-2) -- (-6,-3.6);
\draw[gray,dotted] (-5,-2) -- (-5,-3.6);

\draw[black, dotted, line width=1mm] (-10.3,-2.8) -- (-10,-2.8);
\draw[black,line width=1mm] (-9.9,-2.8) -- (-9,-2.8);
\draw[black,line width=1mm] (-9,-2) -- (-8,-2);
\draw[black,line width=1mm] (-8,-2.8) -- (-7,-2.8);
\draw[black,line width=1mm] (-7,-2) -- (-6,-2);
\draw[black,line width=1mm] (-6,-2.8) -- (-5,-2.8);
\draw[blue,line width=1mm] (-5,-2) -- (-4,-2);
\draw[blue, dotted,line width=1mm] (-4,-2) -- (-3.4,-2);

\draw[black,->] (-2.9,-2.8) -- (-1.6,-2.8) node[midway, below, sloped] {{\small $w_1 \to w_0$}};

\draw[black,thin] (-0.6,-2) -- (4,-2);
\draw[black,thin] (-0,-2.8) -- (4.5,-2.8);
\draw[black,->] (-0.7,-3.6) -- (4.5,-3.6);

\draw (-1,-2) node{$t$};
\draw (-1,-2.8) node{$\phi$};
\draw (4.3,-3.8) node{$z$};

\draw[gray,dotted] (1,-2) -- (1,-3.6);
\draw[gray,dotted] (2,-2) -- (2,-3.6);
\draw[gray,dotted] (3,-2) -- (3,-3.6);

\draw (-0.2,-3.9) node{{\small $(d)$}};
\draw (1,-3.8) node{{\small $w_2$}};
\draw (2,-3.8) node{{\small $w_3$}};
\draw (3,-3.8) node{{\small $w_4$}};

\draw[black, dotted, line width=1mm] (-0.6,-2.8) -- (-0.2,-2.8);
\draw[black,line width=1mm] (-0.1,-2.8) -- (1,-2.8);
\draw[black,line width=1mm] (1,-2) -- (2,-2);
\draw[black,line width=1mm] (2,-2.8) -- (3,-2.8);
\draw[blue,line width=1mm] (3,-2) -- (4,-2);
\draw[blue, dotted,line width=1mm] (4.1,-2) -- (4.5,-2);

\draw[black,->] (2,-0.5) -- (2,-1.5);
\draw (3,-1) node{{\small $w_1\to -\infty$}};

\draw (2,-4.5) node{{ {\it One accelerating black hole}}};

\end{tikzpicture}
\caption{{\small $(a)$ Rod diagram for a collinear three-black hole system (an Israel-Khan solution).
The limit $w_0 \to -\infty$ and $w_5 \to \infty$ gives $(b)$ the single black hole in the expanding bubble. When sending $w_5\to\infty$ in $(a)$ we obtain
$(c)$ two accelerating black holes. Collapsing one timelike rod in the latter gives
$(d)$ the C-metric for a single accelerating black hole. The rod limits commute, so from the two accelerating black holes in $(c)$, the limit $w_0\to-\infty$ gives $(b)$ a single black hole in a bubble.}}
\label{fig:3bh--1bh+bubble}
\end{figure}
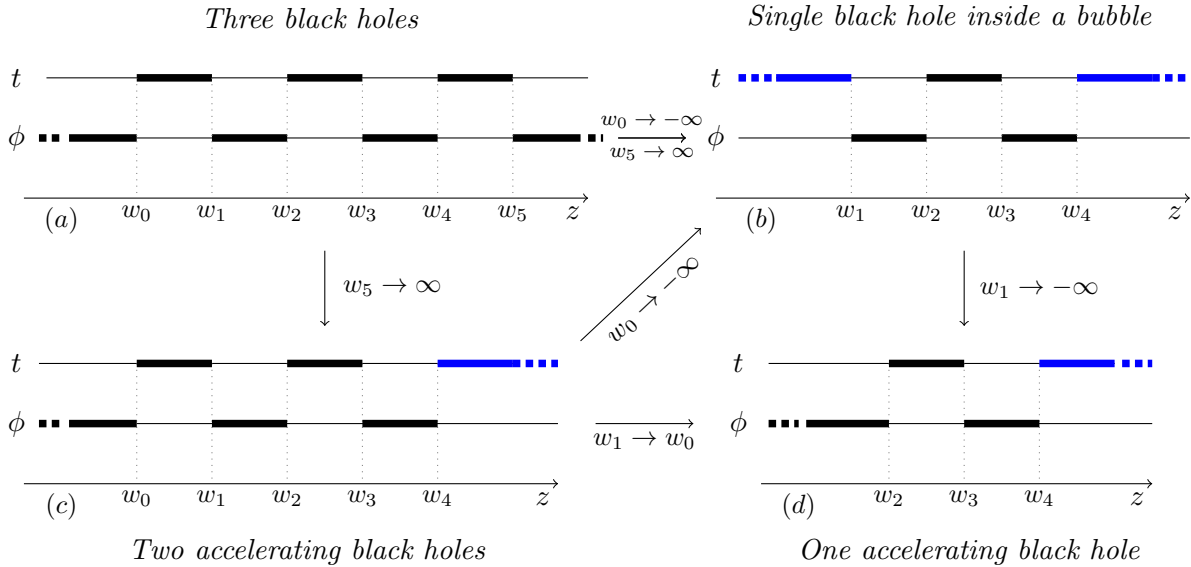

\paragraph{4D.}
One obvious extension of the binary system studied above is the three-black hole configuration contained in the Israel-Khan solution and represented in Fig.~\ref{fig:3bh--1bh+bubble}(a).
To get the Schwarzschild black hole inside the bubble of nothing, we extend the peripheral timelike rods to infinity, taking the limits $w\to-\infty$ and $w_5\to\infty$.

From the black hole in the bubble we can also generate a metric describing a point-like Curzon-Chazy particle embedded in the bubble.
The procedure is similar to the one used to obtain the Bonnor-Swaminarayan solution from an accelerating binary black hole system~\cite{Astorino:2021rdg}.

Moreover it is very clear, in the 4D setting, how to generate accelerating black hole metrics from the black holes in the bubble, for any number of collinear black holes.
It is sufficient to push away only one of the two poles defining the bubble, for instance $w_0 \to -\infty$ in the binary configuration of Sec.~\ref{sec:binary+bubble}.
In Fig.~\ref{fig:3bh--1bh+bubble} we picture the single black hole case. The limiting process, however, introduces irremovable conical singularities, unless an external background field is introduced, as in~\cite{Astorino:2021rdg}.
It is clear that this procedure cannot be pursued in 5D. In this case, there is only a single rod determining the bubble horizon, but more importantly, the five-dimensional C-metric for a uniformly accelerating black hole would have different symmetry ($SO(3)$ rotations, rather than $U(1)^2$) and not be in the Weyl class of solutions.

\begin{figure}[t]
\centering
\begin{tikzpicture}

\draw (-7,2.5) node{{{\it Black di-ring}}};

\draw[black,thin] (-10.2,1.8) -- (-3.6,1.8);
\draw[black,thin] (-9.5,1) -- (-3.6,1);
\draw[black,thin] (-10.2,0.2) -- (-4,0.2);

\draw[black, dotted, line width=1mm] (-10.3,1) -- (-10,1);
\draw[black,line width=1mm] (-9.9,1) -- (-9,1);
\draw[black,line width=1mm] (-9,1.8) -- (-8,1.8);
\draw[black,line width=1mm] (-8,1) -- (-7,1);
\draw[black,line width=1mm] (-7,1.8) -- (-6,1.8);
\draw[black,line width=1mm] (-6,1) -- (-5,1);
\draw[black,line width=1mm] (-5,0.2) -- (-4,0.2);
\draw[black, dotted, line width=1mm] (-4,0.2) -- (-3.5,0.2);

\draw[gray,dotted] (-9,1.8) -- (-9,-0.6);
\draw[gray,dotted] (-8,1.8) -- (-8,-0.6);
\draw[gray,dotted] (-7,1.8) -- (-7,-0.6);
\draw[gray,dotted] (-6,1.8) -- (-6,-0.6);
\draw[gray,dotted] (-5,1.8) -- (-5,-0.6);

\draw (-10.1,-0.9) node{{\small $(a)$}};
\draw (-9,-0.8) node{{\small $w_1$}};
\draw (-8,-0.8) node{{\small $w_2$}};
\draw (-7,-0.8) node{{\small $w_3$}};
\draw (-6,-0.8) node{{\small $w_4$}};
\draw (-5,-0.8) node{{\small $w_5$}};

\draw (-3.5,-0.8) node{$z$};

\draw (-10.6,1.8) node{$t$};
\draw (-10.6,1) node{$\phi$};
\draw (-10.6,0.2) node{$\psi$};

\draw[black,->] (-10.5,-0.6) -- (-3.5,-0.6);


\draw[black,->] (-3.1,0.6) -- (-1.7,0.6) node[midway, sloped, below] {{\small $w_1\to -\infty$}};


\draw (1.8,2.5) node{{ {\it Black ring inside a bubble}}};
\draw[black,thin] (-0.5,1.8) -- (5,1.8);
\draw[black,thin] (-0.9,1) -- (5,1);
\draw[black,thin] (-0.9,0.2) -- (4.5,0.2);

\draw[blue, dotted, line width=1mm] (-0.9,1.8) -- (-0.5,1.8);
\draw[blue,line width=1mm] (-0.5,1.8) -- (0.5,1.8);
\draw[black,line width=1mm] (0.5,1) -- (1.5,1);
\draw[black,line width=1mm] (1.5,1.8) -- (2.5,1.8);
\draw[black,line width=1mm] (2.5,1) -- (3.5,1);
\draw[black,line width=1mm] (3.5,0.2) -- (4.5,0.2);
\draw[black, dotted, line width=1mm] (4.5,0.2) -- (5,0.2);

\draw[gray,dotted] (0.5,1.8) -- (0.5,-0.6);
\draw[gray,dotted] (1.5,1.8) -- (1.5,-0.6);
\draw[gray,dotted] (2.5,1.8) -- (2.5,-0.6);
\draw[gray,dotted] (3.5,1.8) -- (3.5,-0.6);

\draw (-0.9,-0.9) node{{\small $(b)$}};
\draw (0.5,-0.8) node{{\small $w_2$}};
\draw (1.5,-0.8) node{{\small $w_3$}};
\draw (2.5,-0.8) node{{\small $w_4$}};
\draw (3.5,-0.8) node{{\small $w_5$}};

\draw (4.8,-0.8) node{$z$};

\draw (-1.3,1.8) node{$t$};
\draw (-1.3,1) node{$\phi$};
\draw (-1.3,0.2) node{$\psi$};

\draw[black,->] (-1.3,-0.6) -- (5,-0.6);


\draw[black,->] (-7,-1.5) -- (-7,-2.5);
\draw (-6,-2) node{{\small $w_5\to w_4$}};


\draw[black,thin] (-10.2,-3.2) -- (-4.5,-3.2);
\draw[black,thin] (-9.5,-4) -- (-4.5,-4);
\draw[black,thin] (-10.2,-4.8) -- (-5,-4.8);

\draw[black, dotted, line width=1mm] (-10.3,-4) -- (-10,-4);
\draw[black,line width=1mm] (-9.9,-4) -- (-9,-4);
\draw[black,line width=1mm] (-9,-3.2) -- (-8,-3.2);
\draw[black,line width=1mm] (-8,-4) -- (-7,-4);
\draw[black,line width=1mm] (-7,-3.2) -- (-6,-3.2);
\draw[black,line width=1mm] (-6,-4.8) -- (-5,-4.8);
\draw[black, dotted, line width=1mm] (-5,-4.8) -- (-4.4,-4.8);

\draw[gray,dotted] (-9,-3.2) -- (-9,-5.6);
\draw[gray,dotted] (-8,-3.2) -- (-8,-5.6);
\draw[gray,dotted] (-7,-3.2) -- (-7,-5.6);
\draw[gray,dotted] (-6,-3.2) -- (-6,-5.6);

\draw (-10.1,-5.9) node{{\small $(c)$}};
\draw (-9,-5.8) node{{\small $w_1$}};
\draw (-8,-5.8) node{{\small $w_2$}};
\draw (-7,-5.8) node{{\small $w_3$}};
\draw (-6,-5.8) node{{\small $w_4$}};

\draw (-4.5,-5.8) node{$z$};

\draw (-10.6,-3.2) node{$t$};
\draw (-10.6,-4) node{$\phi$};
\draw (-10.6,-4.8) node{$\psi$};

\draw[black,->] (-10.5,-5.6) -- (-4.4,-5.6);

\draw (-7.1,-6.5) node{{ {\it Black Saturn}}};


\draw[black,->] (-3.4,-4.2) -- (-1.9,-4.2) node[midway, sloped, below] {{\small $w_1\to -\infty$}};


\draw[black,thin] (-0.5,-3.2) -- (4,-3.2);
\draw[black,thin] (-0.9,-4) -- (4,-4);
\draw[black,thin] (-0.9,-4.8) -- (3.5,-4.8);

\draw[blue, dotted, line width=1mm] (-0.9,-3.2) -- (-0.5,-3.2);
\draw[blue,line width=1mm] (-0.5,-3.2) -- (0.5,-3.2);
\draw[black,line width=1mm] (0.5,-4) -- (1.5,-4);
\draw[black,line width=1mm] (1.5,-3.2) -- (2.5,-3.2);
\draw[black,line width=1mm] (2.5,-4.8) -- (3.5,-4.8);
\draw[black, dotted, line width=1mm] (3.6,-4.8) -- (4.1,-4.8);

\draw[gray,dotted] (0.5,-3.2) -- (0.5,-5.6);
\draw[gray,dotted] (1.5,-3.2) -- (1.5,-5.6);
\draw[gray,dotted] (2.5,-3.2) -- (2.5,-5.6);

\draw (-0.9,-5.9) node{{\small $(d)$}};
\draw (0.5,-5.8) node{{\small $w_2$}};
\draw (1.5,-5.8) node{{\small $w_3$}};
\draw (2.5,-5.8) node{{\small $w_4$}};

\draw (3.9,-5.8) node{$z$};

\draw (-1.3,-3.2) node{$t$};
\draw (-1.3,-4) node{$\phi$};
\draw (-1.3,-4.8) node{$\psi$};

\draw[black,->] (-1.3,-5.6) -- (4.1,-5.6);

\draw (1.8,-6.5) node{{ {\it Black hole inside a bubble}}};

\draw[black,->] (1.5,-1.5) -- (1.5,-2.5);
\draw (2.4,-2) node{{\small $w_5\to w_4$}};

\end{tikzpicture}
\caption{{\small $(a)$ Rod diagram for a coaxial double black ring system.
The limit $w_1 \to -\infty$ gives $(b)$ the single black ring in the expanding bubble.
The limit of $(a)$ for $w_5\to w_4$ gives $(c)$ the black Saturn.
Its limit for $w_1 \to -\infty$ gives $(d)$ the five-dimensional black hole in an expanding bubble.
The rod limits commute, so the latter diagram can also be obtained from $(b)$ for $w_5\to w_4$.}}
\label{fig:4diagrams}
\end{figure}
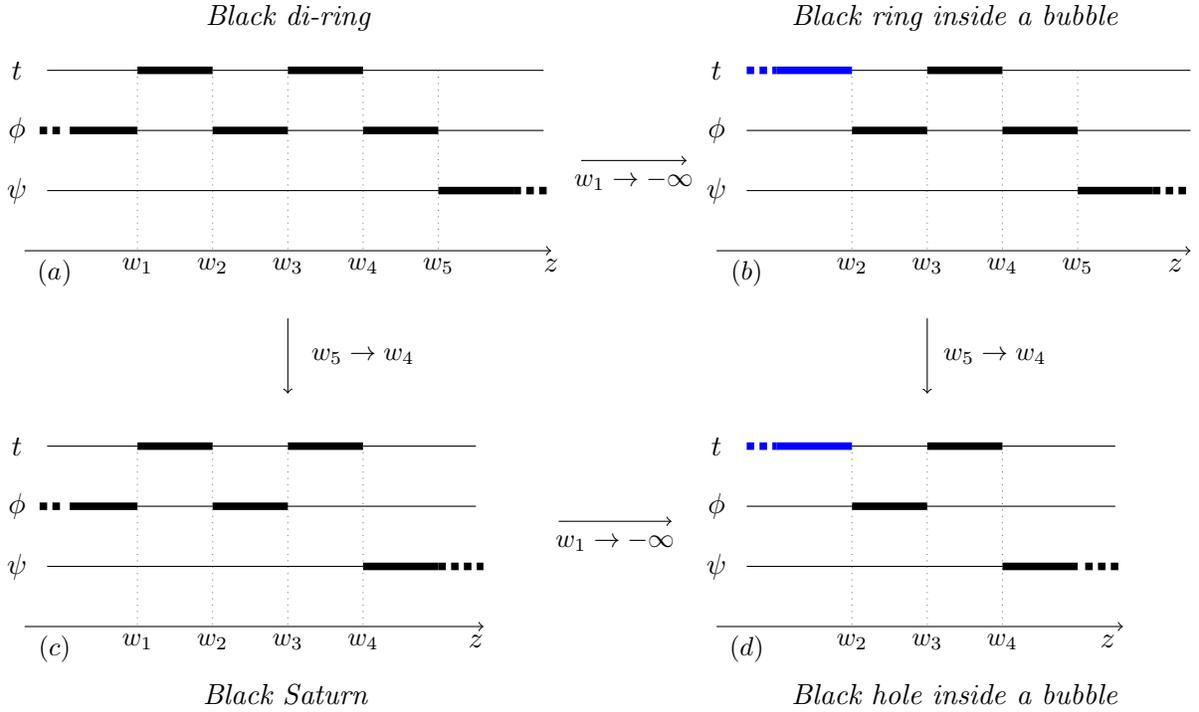

\paragraph{5D.}
It is interesting that all the five-dimensional configurations studied in this paper can be obtained by performing limits in the black di-ring configuration of Fig.~\ref{fig:4diagrams}(a).

For instance, to recover the black ring-bubble of nothing of Fig.~\ref{fig:4diagrams}(b) (which also corresponds to Fig.~\ref{fig:ring-bubble}), we simply send $w_1\to-\infty$ in the black di-ring diagram.
Furthermore, one can also obtain the black hole inside the bubble from the latter by taking $w_5\to w_4$ to remove a spacelike finite rod.

On the other hand, if we take $w_5\to w_4$ in the di-ring diagram, we recover the black Saturn~\cite{Elvang:2007rd} of Fig.~\ref{fig:4diagrams}(c).
From this diagram, we can send $w_1\to-\infty$ to obtain the 5D black hole-bubble of nothing, which corresponds to the superposition of the two diagrams of Fig.~\ref{fig:5D-1bh-1bubble}.

\section{Discussion and outlook}\label{sec:discuss}

Black holes and bubbles of nothing are some of the most elementary solutions in General Relativity, and in this article we have argued that their properties are closely interrelated. By revealing how bubbles are present in black hole systems, we have learned that the spacetime expansion in the bubble is driven by the same phenomenon that makes the volume inside a black hole grow.

The basic idea is simple enough to lend itself to easy generalization. Whenever a small gap region appears between black hole horizons in a maximally extended geometry, it will contain an expanding bubble of nothing. The bubble is a minimal cycle that links the Einstein-Rosen bridges of the system. In the simplest instance, namely, the four-dimensional black hole binary in Sec.~\ref{subsec:bbhbub}, the topology of a Cauchy slice is $S^1\times S^2-\{0\}$ (the point at infinity is removed), and the bubble is the minimal $S^1$ in it. Similarly, for the black ring, the spatial topology is $S^2\times S^2-\{0\}$, and the bubble is the minimal $S^2$. In the more general `ringoids' of \cite{Kleihaus:2009wh,Kleihaus:2010pr} with spatial topology $S^{d-3}\times S^2-\{0\}$ we find $S^{d-3}$ bubbles. We have even considered more complex bubbles topologies, such as the double $S^2$ bubble in Sec.~\ref{subsec:5dbbh}, and we have identified that a collinear black hole binary in $d$ dimensions, with Cauchy slices that are $S^1\times S^{d-2}-\{0\}$, must have an $S^1$ bubble.

Such solutions for binary black holes are not known explicitly in arbitrary dimensions, but we can easily find configurations with two disconnected horizons which can be regarded, in the sense explained above, as possessing expanding bubbles of nothing. The Schwarzschild-de Sitter solution
\begin{equation}
    ds^2=-\left( 1-\frac{\mu}{r^{d-3}}-\frac{r^2}{L^2}\right) +\frac{dr^2}{1-\frac{\mu}{r^{d-3}}-\frac{r^2}{L^2}} +r^2 d\Omega_{d-2}
\end{equation}
with
\begin{equation}
    0<\mu<\mu_N\equiv\frac{2}{d-3}\left(\frac{d-3}{d-1}\right)^\frac{d-1}{2} L^{d-3}
\end{equation}
has a cosmological horizon and a black hole horizon. In the maximal analytic extension (and identifying regions beyond the horizons to form a spatial circle) the spatial sections have topology $S^1\times S^{d-2}$. The $S^1$ expands in time, inside the black hole and in the time-dependent region beyond the de Sitter horizon. The analogue of the bubble limit is the limit $\mu\to \mu_N$, in which (after rescaling $t$) we recover the Nariai solution
\begin{equation}
    ds^2=\frac{L^2}{d-1}\left[-(1-\xi^2)dt^2 +\frac{d\xi^2}{1-\xi^2}+(d-3) d\Omega_{d-2}\right]
\end{equation}
with horizons at $\xi=\pm 1$. We can change coordinates in this metric (see footnote~\ref{foot:change}) to the form
\begin{equation}
    ds^2=\frac{L^2}{d-1}\left[-dT^2 +\cosh^2 T d\chi^2 +(d-3) d\Omega_{d-2}\right]
\end{equation}
with $0\leq \chi\leq 2\pi$. Here we recognize the essential features of the $n=1$ circular bubble \eqref{bon1}, only that now the inhomogeneous, non-compact $(r,\phi)$ cigar is replaced by a round $S^{d-2}$. Thus, bubble-of-nothing-like expansion is indeed pervasive and connected in wide generality to the phenomenon of spacetime expansion.

It is now clear, given the plethora of black hole topologies and multi-black hole configurations that are possible in higher dimensions \cite{Emparan:2009vd}, that we can expect a large variety of expanding bubbles of nothing, even in vacuum gravity. Many of them are unlikely to admit a closed exact solution, but it is intuitively useful to first conceive of them as black hole configurations, as this helps identify new possibilities. 
It would be interesting to know how general the converse is, that is, whether for any expanding bubble one can find a black hole configuration that contains it as a limit.

We have also proven, with explicit examples, that the bubble expansion acts on gravitating systems in much the same way as de Sitter-type inflation: It counteracts the gravitational attraction between localized objects and allows novel static multi-black hole configurations. Again, this phenomenon is likely valid for all the expanding bubbles that we have mentioned above, and more generally for other bubbles.
 
Our arguments show that the expanding bubble of nothing is already present in the binary or black ring \emph{even before taking the small-gap limit}, in the sense that there is a minimal cycle that links the system of Einstein-Rosen bridges and which expands because it stretches inside the black holes. Taking the small-gap limit makes the bubble more symmetric and uniform, and its expansion becomes asymptotically uniform and eternal, since in the limit the interior black hole singularity is pushed away to infinitely late time. If the black hole system were of finite size, or if it were to merge or collapse, the duration of the expansion would instead be limited, ending on a singularity. But expanding bubbles of nothing, in the above sense, seem pervasive in black hole systems with multiple or non-spherical horizons.\footnote{However, this does not mean that they must admit a good limit to a bubble solution; we already mentioned in Sec.~\ref{subsec:ringtobubble} that the equilibrium rotating black ring does not admit it, even though at any finite radius it has a bubble in the sense explained above.} 

Does this mean that we should expect bubbles of nothing in astrophysical, dynamical binaries more realistic than the static ones we have studied? Unfortunately, the answer is no. The topology of a binary where the black holes formed from collapsing matter is different than in the maximal analytic extensions we have considered. Collapsing black hole geometries do not have bifurcation surfaces nor Einstein-Rosen bridges. Even though space expands inside a collapsing black hole, the topology of the Cauchy slices is trivial, and these binaries will not contain any minimal cycle.

However, even if expanding bubbles of nothing may not be present in the sky above, their connection to more conventional black hole systems provides a new, illuminating perspective on their properties and makes them seem more accessible. Since they behave in many ways like de Sitter space---but without a cosmological constant, and with non-compact horizons---they may provide new venues in which to investigate the holographic description of expanding spacetime, possibly exploiting their relation to Einstein-Rosen bridges and the interiors of black hole systems.


\section*{Acknowledgements}

RE thanks Benson Way for useful conversations. MA and AV are partly supported by MIUR-PRIN contract 2017CC72MK003.
RE acknowledges financial support from MICINN grant PID2019-105614GB-C22, AGAUR grant 2017-SGR 754, and State Research Agency of MICINN through the ``Unit of Excellence Maria de Maeztu 2020-2023" award to the Institute of Cosmos Sciences (CEX2019-000918-M).


\appendix

\section{Another limit to the 5D bubble}
\label{app:otherring}

The static black ring can be written in the coordinates introduced in \cite{Emparan:2004wy} in the form
\begin{equation}
ds^2=-\frac{F(y)}{F(x)}d\tilde{t}^2 +\frac{R^2}{(x-y)^2}F(x)\left[ (y^2-1)d\tilde{\psi}^2 +\frac1{F(y)}\frac{dy^2}{y^2-1}+\frac1{F(x)}\frac{dx^2}{1-x^2}+(1-x^2)d\tilde{\phi}^2\right]\,,
\end{equation}
where now
\begin{equation}
F(\xi)=1+\nu\xi\,,
\end{equation}
with $0\leq \nu <1$. In the rotating metric in~\cite{Emparan:2004wy}, we have set $\lambda=\nu$ to obtain a static solution.
The ring becomes fatter as $\nu\to 1$.
Then, the bubble of nothing~\eqref{5dbubble} is obtained taking the limit $\epsilon\to 0$ with
\begin{equation}
\nu = 1-\epsilon \,, \qquad
y = -1-\epsilon\, \xi^2\,, \qquad 
x = -1+\frac{2r_0^2}{r^2} \,,
\end{equation}
where $r_0=R$, and rescaling
\begin{equation}
\tilde{t} = r_0\sqrt{\frac{2}{\epsilon}}\,t\,,\qquad 
\tilde{\psi} = \frac{1}{\sqrt{\epsilon}}\,\psi\,,\qquad
\tilde{\phi} = \frac{1}{\sqrt{2}}\phi\,.
\end{equation}

\end{document}